\documentclass[12pt,a4paper]{article}
\usepackage[utf8]{inputenc}
\usepackage{natbib}
\usepackage[top=26truemm,bottom=26truemm,left=26truemm,right=26truemm]{geometry}

\usepackage{setspace}

\usepackage[mathlines]{lineno}
\usepackage{physics}
\usepackage{enumerate}

\usepackage{mathtools}
\mathtoolsset{showonlyrefs=true} 
\allowdisplaybreaks

\usepackage{booktabs} 
\usepackage{amsthm, amssymb, amsmath, appendix}
\usepackage{hyperref}

\theoremstyle{definition}
\newtheorem{definition}{Definition}
\newtheorem{remark}{Remark}

\theoremstyle{plain}
\newtheorem{theorem}{Theorem}
\newtheorem{subtheorem}{Theorem}[theorem]
\newtheorem{lem}{Lemma}
\newtheorem{prop}{Proposition}

\newtheorem{axiom}{Axiom}
\newtheorem{axiomB}{Axiom$^\ast$}
\newtheorem*{amatheorem}{Amarante's Representation Theorem}
\newtheorem*{gmmtheorem}{GMM's Representation Theorem}

\newcommand{\stsucc}{\succ\!\!\!\succ}

\usepackage{tikz}
\usetikzlibrary{intersections, calc, arrows.meta}
\usepackage{cancel}

\title{Cautious Dual-Self Expected Utility and \\ Weak Uncertainty Aversion\thanks{The authors are grateful to Victor Aguiar, Yutaro Akita, Soo Hong Chew, Spyros Galanis, Youichiro Higashi, Ryota Iijima, Shaowei Ke, Nobuo Koida, Fabio Maccheroni, Satoshi Nakada, Daisuke Nakajima, Kemal Ozbek, Kota Saito, Taishi Sassono, Koichi Tadenuma, Seiji Takanashi, Norio Takeoka, Takashi Ui, Yusuke Yamaguchi, and Songfa Zhong for their helpful comments. The authors would like to thank participants at the Japanese Economic Association 2025 Autumn Meeting (Chukyo University), Risk, Uncertainty, \& Decision (the University of Manchester), Decision Theory Workshop (Nagasaki University), Social Choice Theory Workshop (the University of Tokyo), Game Theory Workshop (Kanazawa), and ``Workshop: Frontiers in Behavioral and Experimental Economics" (Hitotsubashi University). 
The financial support from JSPS KAKENHI (No.23KJ0979 \& 25KJ1298) and Joint Usage and Research Center Programs of Hitotsubashi University (No.IERPK2517) is gratefully acknowledged.
}}
\author{Kensei Nakamura\thanks{Graduate School of Economics, Hitotsubashi University, Kunitachi, Tokyo 186-8601, Japan. E-mail: kensei.nakamura.econ@gmail.com}\hspace{1mm} and Shohei Yanagita\thanks{Graduate School of Economics, Hitotsubashi University, Kunitachi, Tokyo 186-8601, Japan. E-mail: shoheiyanagita@gmail.com}}
\date{\today}

\begin{document}

\vspace{-5mm}
\maketitle
\vspace{-3mm}

\onehalfspacing

\begin{abstract}
    \citeauthor{gilboa1989maxmin}'s (\citeyear{gilboa1989maxmin}) uncertainty aversion plays a central role in decision theory and economics, yet many inconsistent behaviors have been observed in experiments. Motivated by this, we study an axiom postulating a minimal degree of uncertainty aversion. Our main result shows that this axiom yields a new class of representations, called cautious dual-self expected utility representations. In this model, two selves in the decision maker's mind play an extensive-form game to determine the belief used for evaluation, where the first mover is selected cautiously. As illustrated by two alternative representations, cautiously choosing between two ``dual'' scenarios is the key implication of our axiom.
    \vspace{0mm}
\end{abstract}

\newpage

\section{Introduction}
\sloppy

As a critique of the subjective expected utility (EU) model, \citet{ellsberg1961risk}  offered a thought experiment indicating that decision makers (henceforth DMs) tend to avoid uncertainty and suggested that the model with a unique prior fails to align with real-world behaviors. 
Following this point, various non-expected utility models  under uncertainty have been proposed and examined in  decision theory.

To capture the uncertainty-averse attitudes of DMs, \citet{gilboa1989maxmin} introduced an axiom called \textit{uncertainty aversion}. This axiom postulates that for any two uncertain prospects such that they are equally desirable to the DM, mixtures of them through randomization are weakly preferred to the original ones. 
Since then, \textit{uncertainty aversion} has played a central role in decision theory, and many models have been proposed and characterized using this axiom, such as the maxmin EU model (\citet*{gilboa1989maxmin}), the variational EU model (\citet*{maccheroni2006ambiguity} (henceforth MMR)), and the uncertainty-averse EU model (\citet*{cerreia2011uncertainty} (henceforth CMMM)).


However, as is often pointed out, individuals in the real world do not always behave as assumed in \textit{uncertainty aversion}. 
Various laboratory experiments have shown that, while subjects usually dislike uncertain situations, the same subjects sometimes seek ambiguity (e.g., \citet{dimmock2015estimating,dimmock2016ambiguity,kocher2018ambiguity,li2018rich}, and for a survey, \citet{trautmann2015ambiguity}).
Since \textit{uncertainty aversion} requires universal uncertainty avoidance, it is not sufficiently flexible to accommodate the various choice patterns observed in the above studies. 

Based on this experimental evidence, we study a DM who exhibits a minimal attitude of uncertainty aversion but does not necessarily satisfy \citeauthor{gilboa1989maxmin}'s original axiom of \textit{uncertainty aversion}.
To reconsider what their axiom postulates, let us begin with an example \textit{à la} Ellsberg's original thought experiment.
Suppose there is an urn containing 100 red, green, blue, or black balls where the number of balls of each color is unknown. Consider two bets that offer a prize depending on the color of the drawn ball: Bet I offers \$100 if the ball is red or green and \$0 otherwise; Bet II offers \$100 if the ball is blue or black and \$0 otherwise (see Table \ref{tab:urn}). A DM regards these two bets as equally preferable.
For these bets, randomization can resolve uncertainty---specifically, the new bet that assigns Bets I and II with probability 1/2 each after drawing a ball is a perfect hedge since it gives the DM a common lottery that yields \$100 with probability 1/2, regardless of the ball's color. 
However, randomization does not always reduce the degree of uncertainty as in the above example. 
For instance, consider another pair of bets: Bet III offers \$105 if the ball is red, \$95 if green, and \$0 otherwise; Bet IV offers \$95 if the ball is red, \$105 if green, and \$0 otherwise. 
The 50-50 randomized bet assigns the lottery that yields \$105 and \$95 with probability 1/2 each if the ball is red or green but \$0 otherwise. 
In this example, it is less clear whether randomization resolves uncertainty compared to the first case, but \textit{uncertainty aversion} postulates preference for randomization even in such cases. 

\begin{table*}
\label{tab:urn}
\centering
\caption{An urn example: Randomization and Uncertainty}
\begin{tabular}{lcccc}
\toprule
& \multicolumn{4}{c}{Color of the Ball} \\
\cmidrule(lr){2-5}
& Red & Green & Blue & Black \\
\midrule
Bet I & \$100 & \$100 & \$0 & \$0 \\
Bet II & \$0 & \$0 & \$100 & \$100 \\
Bet III & \$105 & \$95 & \$0 & \$0 \\
Bet IV & \$95 & \$105 & \$0 & \$0 \\
\bottomrule
\end{tabular}
\end{table*}

Motivated by these observations, we consider an axiom called \textit{weak uncertainty aversion}
It requires that a DM exhibits an uncertainty-averse attitude only when subjective uncertainty is reduced to objective risk thorough randomization, as in the first example. 
In other words, \textit{weak uncertainty aversion} postulates that for two indifferent uncertain prospects, a DM prefers randomization between them if it constitutes a perfect hedge. 
This axiom still captures the Ellsberg-type choice patterns and is consistent with the evidence from the laboratory experiments mentioned above.\footnote{Note that \textit{weak uncertainty aversion} is closely related to ambiguity seeking for small-likelihood events, which has been observed in many experiments. 
For a detailed discussion, see Example 1 of \citet{chandrasekher2022dual}.}

The objective of this paper is to clarify the implications of \textit{weak uncertainty aversion} in a class of preferences with a basic structure, known as \textit{invariant biseparable preferences} (\citet*{ghirardato2004differentiating} (henceforth GMM)). 
More specifically, we examine weak orders that satisfy \textit{weak uncertainty aversion} together with \textit{non-triviality}, \textit{continuity}, \textit{monotonicity}, and \textit{certainty independence}---thus, we study what preferences can be accommodated if we impose \textit{weak uncertainty aversion} instead of \textit{uncertainty aversion} in \citeauthor{gilboa1989maxmin}'s characterization of the maxmin EU model.
The main result of this paper shows that DMs with these preferences evaluate each (Anscombe–Aumann) act $f$ according to 
\begin{equation}
    \label{eq:cautious_dual}
        U(f) = \min \qty{ \max_{P \in \mathbb{P}} \min_{p\in P} \mathbb{E}_p [u(f)], \min_{P \in \mathbb{P}} \max_{p\in P} \mathbb{E}_p [u(f)]}, 
    \end{equation}
where $\mathbb{P}$ is a collection of subsets of probability distributions over states. 

To explain the interpretation of \eqref{eq:cautious_dual}, we briefly discuss a result of \citet*{chandrasekher2022dual} (henceforth CFIL). 
They showed that invariant biseparable preferences admit dual-self EU representations: That is, a DM with an invariant biseparable preference can be viewed as evaluating each act $f$ according to  
\begin{equation}
\label{eq:dualEU}
        U(f) = \max_{P \in \mathbb{P}} \min_{p\in P} \mathbb{E}_p [u(f)], 
\end{equation}
where $\mathbb{P}$ is a collection of subsets of probability distributions over states, as in \eqref{eq:cautious_dual}. 
The maximization stage represents the DM's uncertainty-seeking attitude, and the minimization stage represents the uncertainty-averse one. 
In \eqref{eq:dualEU}, the belief $p$ selected when evaluating $f$ can be deemed as the outcome of an extensive-form game played by two selves in the DM's mind, \textit{Optimism} and \textit{Pessimism}: 
First, Optimism chooses a subset $P$ of probabilities over the state space from $\mathbb{P}$ aiming to maximize the DM's expected utility of $f$, and then Pessimism chooses a probability $p$ from $P$ aiming to minimize the expected utility of $f$. 
The parameter $\mathbb{P}$ represents how much influence each of the two selves has on determining the belief used for evaluation. 

In the dual-self EU model \eqref{eq:dualEU}, the DM considers only the game where Optimism takes an action first.\footnote{Note that the dual-self EU preference \eqref{eq:dualEU} can be represented by a min-of-max form with another collection $\mathbb{Q}$, which is different from $\mathbb{P}$ in general. Therefore, once fixing a collection representing a preference, only one of the extensive-form games is considered. \label{fn_dseu}}
Similarly, its ``dual" game---that is, the game where Pessimism first chooses a subset $P$ from $\mathbb{P}$ and then Optimism chooses a probability $p$ from $P$---can be considered.
In our representation \eqref{eq:cautious_dual}, when evaluating each act, the DM takes both scenarios into account and then adopts the game that yields the lower expected utility in a cautious manner. 
Thus, we refer to these representations as \textbf{\textit{cautious dual-self EU representations}}. 
Our main theorem shows that \textit{weak uncertainty aversion} leads to the representations with these three-layer intrapersonal belief-selection games.

In the literature, other than the dual-self EU representations, two more ways of representing invariant biseparable preferences have been proposed. 
One is a generalized version of the $\alpha$-maxmin EU model due to GMM. 
The other one is the model proposed by \citet{amarante2009foundations}, which evaluates each act by the Choquet integral of the function that assigns expected utility to each belief with respect to some capacity. 
We examine the implications of \textit{weak uncertainty aversion} based on these representations as well. 
Our results show that as in cautious dual-self EU representations, the DMs with these preferences evaluate each act \textit{by considering two ``dual" scenarios and adopting the worse one in a cautious manner}. 
That is, from any of the three representations of invariant biseparable preferences, we can obtain a common structure using \textit{weak uncertainty aversion}.

It should be noted that \textit{weak uncertainty aversion} itself is not a new concept. \citet{chateauneuf2002diversification}  introduced a similar axiom, which considers randomization among any finite number of indifferent acts, and investigated its implications under the Choquet EU model. \citet{siniscalchi2009vector} introduced a more restricted axiom of uncertainty aversion than ours. It requires an uncertainty-averse attitude if the half mixture of the original acts is a perfect hedge.\footnote{This difference becomes significant when considering more general preferences than the invariant biseparable preferences. For details, see Section \ref{sec_disc}.}  
Furthermore, \citet{aouani2021propensity} and CFIL studied \textit{weak uncertainty aversion}. 
\citet{aouani2021propensity} characterized the property of capacities in the Choquet EU preferences using \textit{weak uncertainty aversion}.  
CFIL showed that if a dual-self EU preference with a collection $\mathbb{P}$ satisfies this axiom, then for any $P, P' \in \mathbb{P}$, $P$ and $P'$ are not disjoint. 
Compared with this result, our theorems provide alternative representations, which clarify an important implication of \textit{weak uncertainty aversion}, for the same class of preferences. 
As these papers, we also provide characterization results of \textit{weak uncertainty aversion} at the level of properties of parameters under the representations of GMM and \citet{amarante2009foundations}. 

Moreover, we offer another justification for the cautious dual-self EU model from a normative perspective. As in \citet*{gilboa2010objective} (henceforth GMMS), we consider the problem of constructing a rational preference from an irrational first criterion (i.e., an incomplete and/or intransitive binary relation).\footnote{Similar models with two binary relations have been considered in \citet{bastianello2022dynamically}, \citet{cerreia2016objective}, \citet{cerreia2020rational}, \cite{faro2019dynamic}, \citet{frick2022objective}, \citet{grant2020objective}, and \citet{kopylov2009choice}.  } 
The first criterion is regarded as  comparisons supported by some objective evidence, and 
the second is interpreted as decision rules for when the DM is forced to choose between alternatives. 
Under this framework, GMMS provided a normative foundation for the maxmin EU model. 
They characterized these preferences as the second criterion, using the first criterion that admits a Bewley representation and axioms about the relationship between the two criteria.
Instead of the Bewley model, we consider a more general class of irrational preferences characterized in Theorem 2 of \citet{lehrer2011justifiable} as the first criterion. 
Our result shows that the cautious dual-self EU preferences can be characterized by two axioms postulating the relationship between the two binary relations, which are modifications of GMMS's axioms.
Since a cautious dual-self EU preference coincides with a maximin EU preference when the first criterion is a Bewley preference, our results can be deemed an extension of the finding in GMMS.

Examining this two-stage model has another merit: It provides a uniqueness result for the parameters in the cautious dual-self EU model.
The uniqueness of parameters is important for conducting comparative statics and identifying the parameters from observed data, but our main characterization does not offer it. 
Since changes in $\mathbb{P}$ of the cautious dual-self EU representation \eqref{eq:cautious_dual} have opposite effects in the max-of-min and the min-of-max part, it is difficult to capture their total effects. 
By introducing the first criterion explicitly, we can avoid this problem and offer some uniqueness result.\footnote{\citet{frick2022objective} also used this strategy to obtain the uniqueness of the parameters in the $\alpha$-maxmin EU model.} 

Finally, we discuss generalizations of  cautious dual-self EU preferences.  
MMR and CMMM studied general models of maxmin EU preferences, and CFIL provided characterization results for their dual-self versions. 
As in the main result, we propose and characterize the cautious dual-self versions of the models proposed by MMR and CMMM. 
We show that these extensions of the cautious dual-self EU model may violate \textit{weak uncertainty aversion} but can be axiomatized using the weaker axiom introduced by \citet{siniscalchi2009vector}.  

This paper is organized as follows: 
Section \ref{sec_main} introduces the formal setup and provides the characterization result of the cautious dual-self EU model. 
Section \ref{sec_other} considers alternative representations based on the results of GMM and \citet{amarante2009foundations}.
Section \ref{sec_rational} presents another justification for the cautious dual-self EU model within the framework introduced by GMMS. 
Section \ref{sec_disc} provides additional results, including characterization results of the generalizations of the cautious dual-self EU model. All proofs are in Appendix.

\section{Cautious dual-self EU representations}
\label{sec_main}

\subsection{Framework}

We consider a model introduced by \citet{anscombe1963definition} and elaborated by \citet{fishburn1970book}. 
Let $S$ be a finite state space and $X$ be the set of lotteries on a set of deterministic prizes.\footnote{
We assume that $S$ is finite since the results by CFIL and \citet{lehrer2011justifiable}, some of which we refer to in our proofs, are based on a finite state space. However, our results are also valid in an infinite state space since their arguments related to our paper can be extended to a general state space. For instance, see Theorem 4.1 of \citet{xia2020decision}, which is a counterpart of Theorem 1 of CFIL in an infinite state space.} 
An \textit{act} is a mapping $f: S \rightarrow X$, and the set of acts is denoted by $\mathcal{F}$.  
With some abuse of notation, we identify an outcome $x\in X$ with the constant act $f$ such that $f(s) = x$ for all $s \in S$. 
We define the mixture operation as follows: For $f,g \in \mathcal{F}$ and $\alpha \in [0,1]$, let $\alpha f + (1 - \alpha) g$ be the act $h$ such that for all $s \in S$, $h(s) =  \alpha f(s) +(1- \alpha) g(s)$. 

A DM has a binary relation $\succsim$ over $\mathcal{F}$. 
For $f,g\in \mathcal{F}$, when we write $f\succsim g$, it means that the DM weakly prefers $f$ to $g$. 
The asymmetric and  symmetric parts of $\succsim$ are denoted by $\succ$ and $\sim$, respectively.

Let $\Delta (S)$ be the set of probability distributions over $S$. We embed $\Delta (S)$ in $\mathbb{R}^S$ and assume that it is endowed with the Euclidean topology. 
We refer to elements of $\Delta (S)$ as beliefs. 
Given $f\in \mathcal{F}$ and  $u : X\rightarrow \mathbb{R}$, let $u(f)$ denote the element of $\mathbb{R}^S$ such that for all $s\in S$, $u(f) (s) = u(f(s))$. 
Furthermore, for $\varphi\in \mathbb{R}^S$, define $\mathbb{E}_p [\varphi] \coloneqq \sum_{s\in S} p(s)\varphi (s)$. 
Let $\mathcal{K} (\Delta (S))$ be the collection of nonempty closed convex subsets of $\Delta (S)$ endowed with the Hausdorff topology. 
We say that $\mathbb{P} \subset\mathcal{K} (\Delta (S))$ is a \textit{belief collection} if it is a nonempty compact collection.

\subsection{Axioms}

We introduce axioms for binary relations over $\mathcal{F}$. 
Binary relations that satisfy the following five axioms are called \textbf{\textit{invariant biseparable preferences}} and have been examined in many papers. 
We omit the detailed explanations for them. 

\begin{axiom}[Non-Triviality]
    For some $f,g\in \mathcal{F}$, $f\succ g$. 
\end{axiom}

\begin{axiom}[Weak Order]
    (i) Completeness: For all $f, g\in \mathcal{F}$, $f\succsim g$ or $g\succsim f$; (ii) Transitivity: For all $f,g,h\in \mathcal{F}$, if $f\succsim g$ and $g\succsim h$, then $f\succsim h$. 
\end{axiom}

\begin{axiom}[Continuity]
    For all $f,g, h\in \mathcal{F}$ with $f\succ g\succ h$, there exist $\alpha, \beta \in (0,1)$ such that $\alpha f + (1 - \alpha ) h \succ g \succ \beta  f + (1 - \beta) h$. 
\end{axiom}

\begin{axiom}[Monotonicity]
    For all $f,g\in \mathcal{F}$, if $f(s) \succsim g(s)$ for all $s\in S$, then $f\succsim g$. 
\end{axiom}

\begin{axiom}[Certainty Independence]
    For all $f,g,\in \mathcal{F}$, $x\in X$ and $\alpha \in (0,1)$, 
    \begin{equation*}
        f \succsim g \iff \alpha f + (1  - \alpha) x \succsim \alpha g + (1  - \alpha) x. 
    \end{equation*}
\end{axiom}

\citet{gilboa1989maxmin} characterized the maxmin EU model using the above five axioms and the axiom of uncertainty aversion. 
Their axiom requires that for any pair of indifferent acts, any mixture of them is weakly preferred to the original ones. 
The formal definition is as follows: 
\begin{axiom}[Uncertainty Aversion]
    For all $f,g\in \mathcal{F}$ and $\alpha\in (0,1)$, if $f\sim g$, then $\alpha f + (1 - \alpha ) g \succsim f$. 
\end{axiom}

This axiom has played a central role in the literature (e.g., MMR; CMMM; \citet{schmeidler1989subjective,epstein2003recursive,chateauneuf2009ambiguity,strzalecki2011axiomatic,saito2015preferences,ke2020randomization}).
However, it has been pointed out that while agents exhibit choice patterns compatible with \textit{uncertainty aversion} in many situations, the same agents sometimes violate it (for a survey, see \citet{trautmann2015ambiguity}).  

Instead of \textit{uncertainty aversion}, we consider a weaker axiom. 
Our axiom requires the uncertainty-averse attitude only when the mixed act smooths out ambiguity and is a perfect hedge. 

\begin{axiom}[Weak Uncertainty Aversion]
    For all $f,g\in \mathcal{F}$ and $\alpha\in (0,1)$, if $f\sim g$ and $\alpha f(s) + (1 - \alpha ) g(s) \sim \alpha f(s') + (1 - \alpha ) g(s') $ for all $s,s' \in S$, then $\alpha f + (1 - \alpha ) g \succsim f$. 
\end{axiom}

Under the additional restriction that $\alpha f(s) + (1 - \alpha ) g(s) \sim \alpha f(s') + (1 - \alpha ) g(s') $ for all $s,s' \in S$, we can deem $\alpha f + (1 - \alpha ) g$ as a constant act. 
This axiom still captures the Ellsberg-type choice patterns and is compatible with the evidence confirmed in laboratory experiments. 

Note that this axiom is not a novel concept.  \citet{chateauneuf2002diversification} introduced a similar axiom, which postulates that for any finite indifferent acts, the mixture of them is preferred to the original acts if the mixed one is a perfect hedge. 
Moreover, CFIL examined the axioms that parameterize the maximum number $k$ of acts that can be used to construct a mixed act. When $k = 2$, it is equivalent to \textit{weak uncertainty aversion}. 
As discussed in Example 1 of CFIL, the case with $k= 2$ is most acceptable, so we focus on this case. 
\citet{siniscalchi2009vector} considered a further weaker axiom than \textit{weak uncertainty aversion} to examine the properties  of parameters in the vector EU model. \citeauthor{siniscalchi2009vector}'s axiom only focuses on the case in which $\alpha = {1\over 2}$. The difference between these two axioms will be discussed in Section \ref{sec_disc}. 
\citet{aouani2021propensity} also examined the implication of \textit{weak uncertainty aversion} in the Choquet EU model.

\subsection{Representation}

Before stating our main characterization theorem, we start with a benchmark result provided in CFIL. 
They showed that the invariant biseparable preferences can be characterized by the the dual-self EU model. 
Formally, we say that a binary relation $\succsim$ over $\mathcal{F}$ admits a \textit{\textbf{dual-self EU representation}} if $\succsim$ is represented by the function $U: \mathcal{F} \to \mathbb{R}$ defined as for all $f\in \mathcal{F}$, 
\begin{equation*}
    U(f) = \max_{P \in \mathbb{P}} \min_{p\in P} \mathbb{E}_p [u(f)], 
\end{equation*}
where  $u:X\rightarrow \mathbb{R}$ is a nonconstant affine function and $\mathbb{P}$ is a belief collection. 
The maximization stage represents the DM's uncertainty-seeking attitude, and the minimization stage corresponds to the uncertainty-averse one. 
This two-stage procedure can be considered a sequential belief-selection game of two selves in the DM's mind, Optimism and Pessimism: 
Given $f\in\mathcal{F}$, Optimism first chooses a subset $P$ of probabilities over the state space from $\mathbb{P}$ aiming to maximize the DM's expected utility of $f$, and then Pessimism chooses a probability $p$ from $P$ aiming to minimize expected utility. 
The parameter $\mathbb{P}$ represents the degree of influence each of the two selves has in determining the chosen belief.

In the dual-self EU model, only the games in which Optimism takes an action first are considered.\footnote{See also Footnote \ref{fn_dseu}.}
However, given a belief collection $\mathbb{P}$, its ``dual" game---that is, the game where Pessimism chooses a subset $P$ from $\mathbb{P}$ first and then Optimism chooses a probability $p$ from $P$---is also plausible. 
We propose a decision-making model that accounts for both games and cautiously chooses one of them.

\begin{definition}
    For a nonconstant affine function $u: X\rightarrow \mathbb{R}$ and a belief collection $\mathbb{P}$, a binary relation $\succsim$ over $\mathcal{F}$ admits a \textit{\textbf{cautious dual-self EU representation}} $(u, \mathbb{P})$ if $\succsim$ is represented by the function $U: \mathcal{F}\to \mathbb{R}$ defined as for all $f\in\mathcal{F}$, 
    \begin{equation}
    \label{eq:def_cautious_dual}
    U(f) = \min \qty{ \max_{P \in \mathbb{P}} \min_{p\in P} \mathbb{E}_p [u(f)], \min_{P \in \mathbb{P}} \max_{p\in P} \mathbb{E}_p [u(f)]
    }. 
\end{equation}
\end{definition}

In the representation \eqref{eq:def_cautious_dual}, the max-of-min part corresponds to the game considered in the original dual-self EU model, and the min-of-max part represents its dual game. 
Thus, when evaluating an act $f$, the DM  considers both scenarios and then adopts the game yielding lower expected utility in a cautious manner. 

Maxmin EU representations are given by the extreme cases in which the set of actions of either the first or second mover is degenerate (i.e., $\mathbb{P}$ is a singleton or each $P\in \mathbb{P}$ is a singleton).  
On the other hand, the maxmax expected utility model is not included except for the case where the prior set is a singleton. 
Furthermore, an $\alpha$-maxmin EU representation $(u, \alpha, P)$, where $\succsim$ is represented by the function 
\begin{equation}
\label{eq:alpha_mm}
    U(f) = \alpha \min_{p \in P} \mathbb{E}_p [u(f)] + (1-\alpha) \max_{p \in P} \mathbb{E}_p [u(f)], 
\end{equation}
is a special case if and only if $\alpha \geq 1/2$ (see Footnote 18 in CFIL and the theorem below). The utility function \eqref{eq:alpha_mm} corresponds to a cautious dual-self EU representation $(u, \mathbb{P})$ with $\mathbb{P} = \{ \alpha P +(1-\alpha) \{ p \} \mid p\in P \}$ or  $\mathbb{P} = \{ \alpha \{ p \} +(1-\alpha) P \mid p\in P \}$. 

Our main theorem states that the invariant biseparable preferences that satisfy \textit{weak uncertainty aversion} can be characterized by the cautious dual-self EU model. 
That is, the cautious way of selecting a game can encapsulate the key implication of \textit{weak uncertainty aversion}. 

\begin{theorem}
\label{thm:cautiousDSEU}
    A binary relation $\succsim$ over $\mathcal{F}$ is an invariant biseparable preference that satisfies \textit{weak uncertainty aversion} if and only if $\succsim$ admits a cautious dual-self EU representation. 
\end{theorem}

Note that because the preferences considered in the above theorem are in the class of invariant biseparable preferences, these preferences can also be represented by the dual-self EU model. 
Indeed, Proposition 3 of CFIL showed that any invariant biseparable preference that satisfies \textit{weak uncertainty aversion} admits a dual-self EU representation $(u, \mathbb{P})$ such that for each $P, P' \in \mathbb{P}$, $P$ and $P'$ are not disjoint. 
Compared with their result, Theorem \ref{thm:cautiousDSEU} provides an alternative representation for the same class of preferences, without any restriction on belief collections. 
As shown in this section and the next, our characterization reveals the essential implication of \textit{weak uncertainty aversion}: 
This weak requirement is closely tied to the cognitive process that evaluates acts in two dual ways and selects the smaller of the two. This second step reflects the DMs' uncertainty-averse attitude, which stems from \textit{weak uncertainty aversion}.

The proof is in Appendix. Instead, we provide here a graphical explanation of why the cautious dual-self EU model can characterize the invariant biseparable preferences with \textit{weak uncertainty aversion}.  

\begin{figure}
    \centering
    \includegraphics[width=0.8\linewidth]{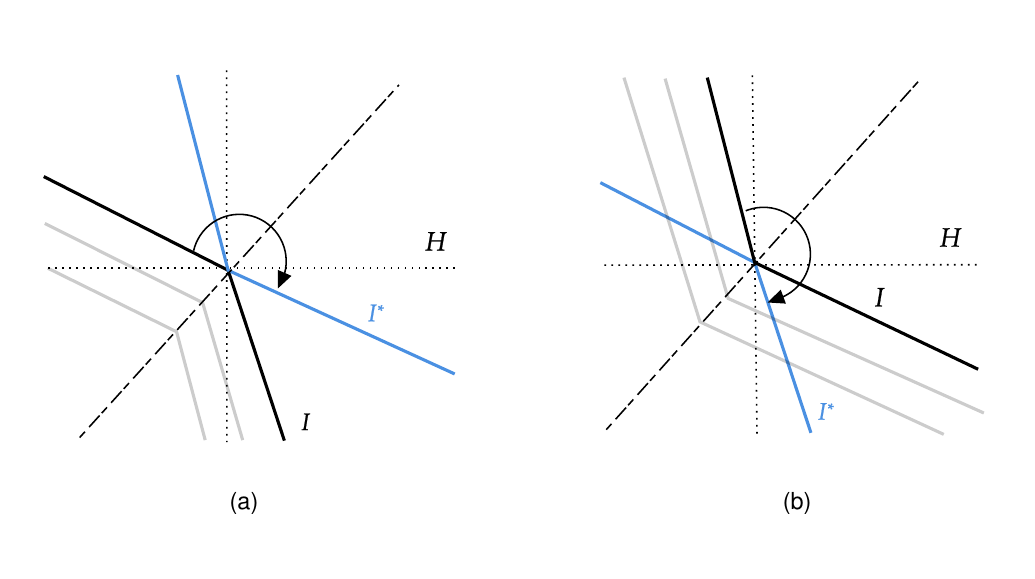}
    \caption{A graphical explanation of the relation between \textit{weak uncertainty aversion} and the cautious dual-self EU model}
    \label{fig_indifference}
\end{figure}

For simplicity, we consider real-valued functions on $\mathbb{R}^S$, where each element represents a utility act.
Note that a pair of acts that can form a perfect hedge is represented by a pair of points in $\mathbb{R}^S$ such that some convex combination of them is in the diagonal line of $\mathbb{R}^S$. 
Thus, under the axioms of invariant biseparable preferences, \textit{weak uncertainty aversion} is equivalent to the condition that for any two-dimensional plane including the diagonal line of $\mathbb{R}^S$, the function restricted to that plane is quasi-concave.\footnote{
\label{fn_cons}For any invariant biseparable preference $\succsim'$, there exist a nonconstant affine function $u': X\to \mathbb{R}$  and a continuous monotone function $I': \mathbb{R}^S \to \mathbb{R}$ such that for all $f,g\in \mathcal{F}$, 
\begin{equation*}
    f\succsim' g\iff I'(u'(f) ) \geq I'(u'(g)). 
\end{equation*}
Furthermore, $I'$ is positively homogeneous and constant-additive. 
(We say that $I'$ is positively homogeneous if for all $\varphi\in \mathbb{R}^S$ and $\alpha > 0$, $I'(\alpha \varphi ) = \alpha I'(\varphi)$; $I'$ is constant-additive if for all $\varphi \in \mathbb{R}^S$ and $\alpha \in \mathbb{R}$, $I'(\varphi + \alpha \mathbf{1}) = I'(\varphi) + \alpha $, where $\mathbf{1} = (1, 1, \cdots, 1)\in\mathbb{R}^S$.) 
} 

Consider the function $I : \mathbb{R}^S\to \mathbb{R}$ such that there exists a belief collection $\mathbb{P}$ with for all $\varphi \in \mathbb{R}^S$, 
\begin{equation*}
    I(\varphi) = \max_{P\in \mathbb{P}} \min_{p\in P} \mathbb{E}_p [\varphi]. 
\end{equation*}
In general, this function is not necessarily quasi-concave on all two-dimensional planes including the diagonal line of $\mathbb{R}^S$, that is, it may violate \textit{weak uncertainty aversion}. 
In the following, we present a general method for constructing a function from $I$ that is compatible with \textit{weak uncertainty aversion}.

Take an arbitrary two-dimensional plane $H$ including the diagonal line.
First, we consider the case where $I$ restricted to $H$ is quasi-convex. 
Then, $I$ is incompatible with \textit{weak uncertainty aversion} on $H$. 
Define the dual function $I^\ast : \mathbb{R}^S\to \mathbb{R}$ of $I$ as for all $\varphi \in \mathbb{R}^S$, 
\begin{equation*}
    I^\ast(\varphi) = \min_{P\in \mathbb{P}} \max_{p\in P} \mathbb{E}_p [\varphi]. 
\end{equation*}
Since the indifference curves of $I^\ast$ can be obtained by rotating those of $I$ by 180 degrees about the origin, $I^\ast$ is quasi-concave (Figure \ref{fig_indifference}(a)).
Therefore, $I^\ast$ is compatible with \textit{weak uncertainty aversion} on $H$. 
Then, consider the function $\widetilde{I}: \mathbb{R}^S\to\mathbb{R}$ defined as $\widetilde{I}  (\varphi) = \min \{  I(\varphi), I^\ast (\varphi)\} $ for all $\varphi\in \mathbb{R}^S$. 
Note that this function corresponds to the cautious dual-self EU representation associated with $\mathbb{P}$. 
Since  $I(\varphi) \geq I^\ast(\varphi)$ holds for each $\varphi \in H$, $\widetilde{I}$ coincides with $I^\ast$ on $H$ and is compatible with  \textit{weak uncertainty aversion}. 

Next, we consider the case where $I$ restricted to $H$ is quasi-concave (Figure \ref{fig_indifference}(b)). 
Then, $I$ on $H$ is compatible with \textit{weak uncertainty aversion}. Furthermore, for each $\varphi \in H$, $I^\ast(\varphi) \geq I(\varphi)$ holds, which implies $\widetilde{I} (\varphi) = I (\varphi)$.  
Therefore, $\widetilde{I}$ on 
$H$ is quasi-concave and compatible with \textit{weak uncertainty aversion} again.

This is why the cautious dual-self EU model can represent the invariant biseparable preferences that satisfy \textit{weak uncertainty aversion}. 

Finally, we provide two remarks on the cautious dual-self EU model: 
\begin{enumerate}[(i)]
\setlength{\itemsep}{0cm}
\setlength{\parskip}{0cm}
\item[(i)] \textbf{Selection of game.} For a cautious dual-self EU representation $(u, \mathbb{P})$, the selected protocol varies  depending on acts in general. However, by the above argument, we can see that for $f, g\in \mathcal{F}$, if there exists $\alpha\in (0,1)$ such that $\alpha f (s)  + (1 -\alpha) g(s) \sim \alpha f(s') + (1-\alpha ) g(s')$ for all $s, s' \in S$, then the selected scenarios when evaluating $f$ and $g$ coincide.\footnote{
We prove this formally. Let $f, g\in \mathcal{F}$ and $\alpha\in (0,1)$ be such that $\alpha f (s)  + (1 -\alpha) g(s) \sim \alpha f(s') + (1-\alpha ) g(s')$ for all $s, s' \in S$. Suppose that the max-of-min scenario is used when evaluating $f$, that is, $\max_{P \in \mathbb{P}} \min_{p\in P} \mathbb{E}_p [u(f)] \leq \min_{P \in \mathbb{P}} \max_{p\in P} \mathbb{E}_p [u(f)]$. Let $x\in X$ be such that $u(x) = \alpha u(f)  + (1 -\alpha) u(g)$, or $u(g)= {1\over 1 - \alpha } u(x)  - {\alpha \over 1 - \alpha } u(f)$. Therefore, 
\begin{align*}
    \max_{P \in \mathbb{P}} \min_{p\in P} \mathbb{E}_p [u(g)] 
    & = \max_{P \in \mathbb{P}} \min_{p\in P} \mathbb{E}_p \qty[{1\over 1 - \alpha } u(x)  - {\alpha \over 1 - \alpha } u(f)]
    = {1\over 1 - \alpha } u(x) -  {\alpha \over 1 - \alpha } \min_{P \in \mathbb{P}} \max_{p\in P} \mathbb{E}_p [ u(f)] \\
    & \leq  {1\over 1 - \alpha } u(x) -  {\alpha \over 1 - \alpha } \max_{P \in \mathbb{P}} \min_{p\in P} \mathbb{E}_p [ u(f)] 
    = \min_{P \in \mathbb{P}} \max_{p\in P} \mathbb{E}_p \qty[{1\over 1 - \alpha } u(x)  - {\alpha \over 1 - \alpha } u(f)] \\
    &  = \min_{P \in \mathbb{P}} \max_{p\in P} \mathbb{E}_p [u(g)],  
\end{align*}
as required. The other case can be shown by a similar argument.
} In other words, if the Optimism-first game (resp. the Pessimism-first game) is used when evaluating $f$, then $g$ is also evaluated using the Optimism-first game (resp. the Pessimism-first game). 

\item [(ii)] \textbf{Single-self interpretation.} As with the dual-self EU model, the cautious dual-self EU model admit a single-self interpretation. 
The max-of-min part represents the utility of a DM who chooses the best prior set by taking an optimal action and then cautiously evaluates acts based on that prior set. The min-of-max part represents the utility of a DM who, after cautiously anticipating which probability set will be available, chooses the best probability distribution by taking an optimal action. 
Hence, the DM with a cautious dual-self EU preference evaluates each act by considering two possible timings of actions and then adopting the less favorable one. 
\end{enumerate}

\section{Alternative representations}
\label{sec_other}

In Theorem \ref{thm:cautiousDSEU}, we have provided a representation theorem for the invariant biseparable preferences satisfying \textit{weak uncertainty aversion} based on the characterization result of CFIL. 
Other than the dual-self EU model, two representations of invariant biseparable preferences were proposed by GMM and \citet{amarante2009foundations}. 
This section offers alternative representations of the preferences considered in Theorem \ref{thm:cautiousDSEU} based on these two papers.
From these results, we find the key implication of \textit{weak uncertainty aversion}: This axiom derives a common structure in which the DM first evaluates an act using two dual scenarios and then adopts the worse one as the final evaluation.

\subsection{Generalized $\alpha$-maxmin EU representations}

We begin with the model provided by GMM. 
They showed that any invariant biseparable preference can be represented by a generalized version of the well-known $\alpha$-maxmin EU model.
To state their result precisely, we introduce several definitions.
For  $f,g\in\mathcal{F}$, we say that $f$ is \textit{unambiguously preferred to} $g$, denoted $f\succsim^\# g$, if for all $h\in\mathcal{F}$ and $\lambda\in(0,1]$, 
\begin{equation*}
    \lambda f+(1-\lambda)h\succsim \lambda g+(1-\lambda)h. 
\end{equation*}
That is, $\succsim^\#$ is the restriction of $\succsim$ that satisfies \textit{independence}.\footnote{For a formal definition, See Axiom \ref{axm:ind}.}
Proposition 5 of GMM showed that if $\succsim$ is an invariant biseparable preference, then  $\succsim^\#$ is represented by a well-known Bewley representation: There exist a nonconstant affine function $u: X\to \mathbb{R}$ and a nonempty closed convex set $P\subset\Delta(S)$ such that for all $f,g\in\mathcal{F}$, 
\begin{equation}
\label{eq:unambiguous}
    f\succsim^\# g \iff \Big[~ \mathbb{E}_p[u(f)]\geq  \mathbb{E}_p[u(g)]~~\text{for all} ~~p\in P ~\Big].
\end{equation}
Thus, $f$ is unambiguously preferred to $g$ if and only if the expected utility of $f$ is higher than that of $g$ for every possible scenario $p\in P$.
By using $\succsim^\#$, we define the relation $\asymp$ over $\mathcal{F}$ as follows:
For any $f,g\in\mathcal{F}$, we write $f\asymp g$ if there exist $x,x'\in X$ and $\lambda,\lambda'\in(0,1]$ such that
\begin{equation*}
    \lambda f+(1-\lambda)x\sim^\# \lambda'g+(1-\lambda')x'.
\end{equation*}
That is, $f \asymp g$ means that $f$ and $g$ possess similar ambiguity in terms of $\succsim^\#$ (i.e., $\succsim$).\footnote{To interpret the relation $\asymp$, it is useful to see  Lemma 8(iii) in GMM.
It states that for all $f,g\in\mathcal{F}$,
\begin{equation*}
    f\asymp g \iff \Big[ \, \forall p,p'\in P, ~~~\mathbb{E}_p[u(f)]\geq \mathbb{E}_{p'}[u(f)] \, \iff\mathbb{E}_p[u(g)]\geq \mathbb{E}_{p'}[u(g)]\Big].
\end{equation*}
Thus, $f\asymp g$ if and only if $f$ and $g$ order possible scenarios identically.}  
Note that  $\asymp$ is an equivalence relation (cf. Lemma 8(ii) of GMM).
For $f\in\mathcal{F}$, we denote by $[f]$ the equivalence class of $\asymp$ containing $f$.

The following is Theorem 11 of GMM.  This states that any invariant biseparable preference can be represented as evaluating each act by an act-dependent weighted sum of the expected utility values in the best and worst scenarios in the DM's mind.

\begin{gmmtheorem}
\label{thm:gmm}
    If $\succsim$ over $\mathcal{F}$ is an invariant biseparable preference, then there exist  a nonconstant  affine function $u:X\rightarrow\mathbb{R}$, a nonempty closed convex set $P\subset\Delta(S)$, and a function $a:\mathcal{F}_{/\asymp}\rightarrow[0,1]$ such that $\succsim$ is represented by the function $U: \mathcal{F}\to \mathbb{R}$ defined as for all $f\in \mathcal{F}$, 
    \begin{equation}
    \label{eq:gen-alpha}
        U(f)=a([f])\min_{p\in P} \mathbb{E}_p [u(f)]+(1-a([f]))\max_{p\in P}\mathbb{E}_p [u(f)].
    \end{equation}
\end{gmmtheorem}

\noindent
We call this representation a \textit{generalized $\alpha$-maxmin EU representation} $(u, P, a)$.
Note that since $a$ is defined on $\mathcal{F}_{/\asymp}$, the weights of $f$ and $g$ coincide if they are in the same equivalence class of $\asymp$.
Moreover, the pair $(u,P)$ in \eqref{eq:gen-alpha} coincides with the one used in the corresponding Bewley representation \eqref{eq:unambiguous} of $\succsim^\#$. 
It should also be emphasized that their result is not an if-and-only-if result: As remarked in GMM, a binary relation represented by the function \eqref{eq:gen-alpha} does not necessarily satisfy \textit{monotonicity}. 

We then analyze what restrictions are imposed on the parameter $\alpha:\mathcal{F}_{/\asymp}\rightarrow[0,1]$ if 
a preference with a generalized $\alpha$-maxmin EU representation satisfies \textit{weak uncertainty aversion}. 
To state our results simply, we introduce additional definitions.
We say that an act $h\in\mathcal{F}$ is \textit{crisp} if $h\asymp x$ for some $x\in X$.
If this relation holds, then $h$ cannot be used for hedging because it exhibits similar ambiguity to the constant act $x$.\footnote{GMM defined crisp acts as follows: $h\in\mathcal{F}$ is \textit{crisp} if for all $f,g\in\mathcal{F}$ and $\lambda\in(0,1)$,
\begin{equation*}
    f\sim g \implies \lambda f+(1-\lambda)h\sim\lambda g+(1-\lambda)h.
\end{equation*}
Their Proposition 10 showed that these two definitions are equivalent.}

For $f\in \mathcal{F}$, let $[f]^\ast \in \mathcal{F}_{/\asymp}$ be an equivalence class such that there exist $g\in [f]^\ast$ and $\alpha \in (0,1)$ with  $\alpha f(s) + (1- \alpha) g (s) \sim \alpha f(s') + (1- \alpha) g (s')$ for all $s,s' \in \mathcal{F}$. 
Note that $[f]^\ast$ is well-defined because of the following lemma. 

\begin{lem}
    Let $\succsim$ over $\mathcal{F}$ be an invariant biseparable preference and $f\in \mathcal{F}$. 
    For all $g,g' \in \mathcal{F}$, if there exist $\alpha, \alpha' \in (0,1)$ with for all $s,s'\in S$, 
    \begin{align*}
        \alpha f(s) + (1- \alpha) g (s) &\sim \alpha f(s') + (1- \alpha) g (s'),  \\
        \alpha' f(s) + (1- \alpha') g' (s) &\sim \alpha' f(s') + (1- \alpha') g' (s'), 
    \end{align*}
    then $[g] = [g']$. 
\end{lem}

Therefore, $[f]^\ast$ is the equivalence class including all acts such that we can construct a complete hedge by mixing $f$ and any of them.

The following result shows that under the generalized $\alpha$-maxmin EU model, \textit{weak uncertainty aversion} can be characterized by simple inequalities about the weights.
These inequalities provide a lower bound on the extent to which the DM accounts for the worst scenario.

\setcounter{theorem}{2}

\begin{subtheorem}
\label{subthm_2a}
    Suppose that a binary relation $\succsim$ over $\mathcal{F}$ admits a generalized $\alpha$-maxmin EU representation $(u, P, a)$ such that $P$ is not a singleton.
    Then, the following statements are equivalent.
    \begin{enumerate}[(i)]
    \setlength{\itemsep}{0cm}
    \setlength{\parskip}{0cm}
        \item [(i)] $\succsim$ satisfies \textit{weak uncertainty aversion}.
        \item [(ii)] For any $f\in\mathcal{F}$ such that $f$ is not crisp, $a([f])+a([f]^\ast)\geq 1$.
    \end{enumerate}
\end{subtheorem}


We exclude the case where $P$ is a singleton since the best and worst scenarios always coincide in this case.
This implies that the function \eqref{eq:gen-alpha} with any function $a: \mathcal{F}_{/\asymp} \to \mathbb{R}$ induces the same binary relation. 

In the $\alpha$-maxmin EU model, it is known that \textit{weak uncertainty aversion} holds if and only if the weight to the worst scenario is weakly greater than $1/2$ (cf. Section 3.1.1 of CFIL). 
Our theorem can be considered a generalization of this result. 

Furthermore, based on the generalized $\alpha$-maxmin EU model, the invariant biseparable preferences that satisfy \textit{weak uncertainty aversion} can be represented in a way similar to the cautious dual-self EU model: 
Instead of considering the two intrapersonal belief-selection games, the DM computes generalized $\alpha$-maxmin expected utility values according to two weight functions. 

\begin{subtheorem}
\label{subthm_2b}
    If a binary relation $\succsim$ over $\mathcal{F}$ is an invariant biseparable preference that satisfies \textit{weak uncertainty aversion}, then there exist  a nonconstant affine function $u:X\rightarrow\mathbb{R}$, a nonempty closed convex set $P\subset\Delta(S)$, and a function $a:\mathcal{F}_{/\asymp}\rightarrow[0,1]$ such that $\succsim$ is represented by
    \begin{equation}
    \label{eq:caudual-genal}
        U(f)=\min\left\{
           \begin{array}{l}
            a([f])\min_{p\in P}\mathbb{E}_p [u(f)]+(1-a([f]))\max_{p\in P}\mathbb{E}_p [u(f)], \\[2mm]
            (1 - a([f]^\ast) ) \min_{p\in P}\mathbb{E}_p [u(f)]+a([f]^\ast) \max_{p\in P}\mathbb{E}_p [u(f)]
           \end{array}
           \right\}.
    \end{equation}
    Furthermore, $\succsim$ represented by the function $U$ satisfies \textit{weak uncertainty aversion}. 
\end{subtheorem}
    
Note that the two weight functions $a([f])$  and $1 - a([f]^\ast) $ can be deemed dual: Indeed, if we set a function $b: \mathcal{F}_{/\asymp}\to[0,1]$ as for all $f\in \mathcal{F}$, $b([f]) = 1 - a([f]^\ast)$, then
\begin{equation*}
    1 - b ([f]^\ast) = 1- (1 - a([f])) = a([f]), 
\end{equation*}
where the first equality follows from $a([f]) = a([f]^{\ast\ast})$.\footnote{Formally, we define $[f]^{\ast\ast} \in \mathcal{F}_{/\asymp}$ as $[f]^{\ast\ast} = [g]^\ast$ for some $g\in [f]^\ast$. } 

Compared with Theorem \ref{thm:cautiousDSEU} and results in CFIL, Theorem \ref{subthm_2a} is the counterpart of Proposition 3 of CFIL since these results characterize the properties of parameters. 
On the other hand, Theorem \ref{subthm_2b} corresponds to Theorem \ref{thm:cautiousDSEU} of this paper:
In both of the representations obtained in Theorems \ref{thm:cautiousDSEU} and \ref{subthm_2b}, the DM first evaluates each act in two dual ways and then chooses the worse one as the evaluation of that act. 

\subsection{Capacity representations}

\citet{amarante2009foundations} provided another representation of invariant biseparable preferences using the Choquet integral. 
Before moving on to our result, we briefly explain \citeauthor{amarante2009foundations}'s characterization result. 

For each $\varphi\in \mathbb{R}^S$, define a function $\kappa_\varphi : \Delta (S) \to \mathbb{R}$ as for all $p\in \Delta (S)$, $\kappa_\varphi (p) = \mathbb{E}_p [\varphi]$.
This transformation assigns to each utility act $\varphi$ a distribution of expected utility values over beliefs. 
We say that a real-valued set function $v$ on $2^{\Delta(S)}$ is a capacity if (i) $v(\emptyset) = 0$, (ii) $v(\Delta (S)) = 1$, and (iii) $v(P) \geq v(P')$ for all $P, P' \subset \Delta (S)$ such that $P\supset P'$. 
For a function $\kappa: \Delta ( S) \to \mathbb{R}$ and a capacity $v$ on $2^{\Delta (S)}$, let 
\begin{equation*}
    \int \kappa  dv = \int^0_{-\infty} \qty{ v(\kappa\geq \beta) -1  } d\beta + \int_0^{\infty}  v(\kappa \geq \beta) d\beta, 
\end{equation*}
where we denote  $v(\{ p \in \Delta (S) \mid  \kappa (p) \geq \beta \}) $ by $v(\kappa  \geq \beta)$ for simplicity. This operator is called the Choquet integral. 
\citeauthor{amarante2009foundations} showed that the invariant biseparable preferences can be represented using these notions. 
The following is a minor modification of Theorem 2 in \citet{amarante2009foundations}. 

\begin{amatheorem}
\label{thm:ama}
    A binary relation $\succsim$ over $\mathcal{F}$ is an invariant biseparable preference if and only if there exist a nonconstant affine function $u:X\rightarrow \mathbb{R}$ and a capacity $v$ on $2^{\Delta (S)}$ such that $\succsim$ is represented by the function $U: \mathcal{F}\to \mathbb{R}$ defined as for all $f\in\mathcal{F}$,
    \begin{equation}
    \label{eq:amarante}
        U(f) = \int\kappa_{u(f)} dv. 
    \end{equation}
\end{amatheorem}

In the original statement of \citeauthor{amarante2009foundations}, for each $\succsim$,  the mapping $\kappa_\varphi$ is defined as a function from $P$ to $\mathbb{R}$ and the capacity $v$ is defined on $2^P$, where $P$ is the subset of $\Delta(S)$ in \eqref{eq:unambiguous}. 
By taking a capacity $v$ on $2^{\Delta(S)}$ appropriately, \citeauthor{amarante2009foundations}'s result can be restated as above.\footnote{
Let $P\subset \Delta (S)$ and $v'$ be a capacity on $2^P$. Define the capacity $v$ on $2^{\Delta(S)}$ as for all $Q\subset \Delta(S)$, $v(Q) = v'(P\cap Q)$. Then, for all affine function $\kappa$ on $\Delta(S)$, 
\begin{equation*}
    \int\kappa dv = \int\kappa dv'. 
\end{equation*}
Therefore, we can take a capacity on $2^{\Delta(S)}$ without loss in \hyperref[thm:ama]{Amarante's Representation Theorem}. 
}

We then see our characterization result. 
For a capacity $v$ on $2^{\Delta(S)}$, define its \textit{dual capacity} $v^\ast$ on $2^{\Delta(S)}$ as for all $P\subset \Delta(S)$, $v^\ast( P ) =  1 - v (P^c)$. 
This notion has been used in cooperative game theory (e.g., \citet{peleg2007introduction}), and several papers in decision theory also considered it (e.g.,  \citet{aouani2021propensity,gul2020calibrated}).
Note that this notion is reflexive in the sense that the double dual capacity of $v$ is $v$ itself.\footnote{The following holds: 
$v^{\ast \ast} (P) = 1 - v^\ast (P^c) = 1- ( 1 - v (P)) = v(P)$. 
}

The following result shows that in \hyperref[thm:ama]{Amarante's Representation Theorem}, \textit{weak uncertainty aversion} can be characterized by a property of a restricted superadditivity of the Choquet integral. 
Furthermore, the invariant biseparable preferences with \textit{weak uncertainty aversion} can be represented using the Choquet integral and the structure derived in the previous theorems: considering the two dual scenarios and adopting the worse one.

\begin{theorem}
\label{thm_ama_weakua}
    Let $\succsim$ be a binary relation over $\mathcal{F}$. The following statements are equivalent: 
    \begin{enumerate}[(i)]
    \setlength{\itemsep}{0cm}
    \setlength{\parskip}{0cm}
        \item [(i)] $\succsim$ is an invariant biseparable preference that satisfies \textit{weak uncertainty aversion}.
        \item [(ii)] For some  nonconstant affine function $u:X\to \mathbb{R}$ and  capacity $v: 2^{\Delta(S)}\to \mathbb{R}$ such that for all affine functions $\kappa$ on $\Delta(S)$, 
        \begin{equation}
        \label{eq:weak_subadd}
            \int \kappa dv + \int - \kappa dv \leq 0, 
        \end{equation}
        $\succsim$ is represented by the function $U: \mathcal{F}\to \mathbb{R}$ defined in \eqref{eq:amarante}. 
        \item [(iii)] There exist  a nonconstant affine function $u: X\to \mathbb{R}$ and a capacity $v$ on $2^{\Delta (S)}$ such that  $\succsim$ is represented by the function $U:\mathcal{F} \to \mathbb{R}$ defined as for all $f \in \mathcal{F}$, 
    \begin{equation*}
        U(f) = \min \qty{ \int \kappa_{u(f)} dv, \int \kappa_{u(f)}  dv^\ast }.
    \end{equation*} 
    \end{enumerate}
\end{theorem}

The second statement is the counterpart of Proposition 3 of CFIL: This characterizes the implication of \textit{weak uncertainty aversion} as a property of the parameter $v$.
The inequality in \eqref{eq:weak_subadd} is a weak version of the superadditivity of the Choquet integral since it can be rewritten as 
\begin{equation*}
    \int \kappa dv + \int - \kappa dv \leq  \int \kappa + (- \kappa) dv ~( = 0). 
\end{equation*}
Superadditivity is closely related to the concavity of the operator, which is in turn related to DMs' uncertainty-averse attitudes. 
Thus, the second statement captures the implication of \textit{weak uncertainty aversion} by restricting the domain in which the condition of superadditivity holds. 

On the other hand, the third statement is the counterpart of Theorem \ref{thm:cautiousDSEU}. 
Under this representation, the DM first computes the value of act $f$ in dual ways based on the original capacity $v$ and its dual capacity $v^\ast$. 
Subsequently, the DM chooses the smaller one as the evaluation of $f$. The second step captures the DM's uncertainty-averse attitude that stems from \textit{weak uncertainty aversion}. 

We have provided several representations of the invariant biseparable preferences satisfying \textit{weak uncertainty aversion}. To summarize, these representations have a common structure, which illuminates the key implication of \textit{weak uncertainty aversion}: evaluating each act in two dual ways and choosing the worse one as the evaluation of that act in a cautious manner. 

\section{Cautious dual-self EU model and rationalization procedure}
\label{sec_rational}

This section provides another justification for the cautious dual-self EU preferences based on the objective/subjective rationality model of GMMS. 
Here, we introduce two binary relations: The first one is a possibly incomplete and intransitive objectively rational preference, and the second one is a complete and transitive subjectively rational preference representing the DM's actual choice. 
The result of this section shows that by imposing normatively appealing axioms not directly related to \textit{weak uncertainty aversion}, the cautious dual-self EU model can be characterized in the second binary relation. 
This result can be interpreted as providing a normative justification for not only the cautious dual-self EU model but also \textit{weak uncertainty aversion}. 
Furthermore, by considering this two-stage model, we obtain a uniqueness result for the parameters in the cautious dual-self EU model.

\subsection{The first criterion and generalized Bewley representations}

We consider two binary relations over $\mathcal{F}$, $\succsim^\ast$ and $\succsim^\land$.
The first relation $\succsim^\ast$ is not necessarily complete and transitive but rational in the sense that it satisfies the independence axiom. 
It can be regarded as representing the DM's objective rationality: For $f,g\in \mathcal{F}$, $f\succsim^\ast g$ means that there is some objective evidence supporting that $f$ is at least as good as $g$. 
Due to a lack of evidence and inconsistencies among the evidence, $\succsim^\ast$ does not always satisfy \textit{completeness} and \textit{transitivity}.\footnote{
Here, we discuss the desirability of dropping  \textit{completeness} and \textit{transitivity}. 
If there is insufficient evidence to compare  two acts, then the DM's first relation $\succsim^\ast$ is naturally incomplete. 
For the violation of \textit{transitivity}, let us consider a collective DM who makes decisions based on advice from experts.  If she adopts majority judgment to aggregate the experts' opinions, then her decision would be intransitive. 
As illustrated by this example, \textit{transitivity} is not an innocent assumption when the DM bases her decisions on multiple sources. 
} 

To deal with a general class of preferences, we impose axioms studied in Theorem 2 of \citet{lehrer2011justifiable} on the first criterion. 
They examined a preference $\succsim$ over $\mathcal{F}$ that satisfy \textit{non-triviality}, \textit{continuity}, and the following axioms:\footnote{More precisely, the continuity axiom in \citet{lehrer2011justifiable} is different from \textit{continuity} defined in this paper. However, we can replace their continuity axiom with ours. } 

\begin{axiom}[Reflexivity]
    For all $f\in \mathcal{F}$, $f\succsim f$. 
\end{axiom}

\begin{axiom}[Completeness for Lotteries]
    For all $x,y\in X$, $x\succsim y$ or $y\succsim x$. 
\end{axiom}

\begin{axiom}[Unambiguous Transitivity]
    For all $f,g, h\in \mathcal{F}$, if either (i) $f(s)\succsim g (s)$ for all $s\in S$ and $g\succsim h$ or (ii) $f\succsim g$ and $g(s) \succsim  h(s)$ for all $s\in S$, then $f\succsim h$. 
\end{axiom}

\begin{axiom}[Independence]
\label{axm:ind}
    For all $f,g,h\in \mathcal{F}$ and $\alpha\in (0,1)$, 
    \begin{equation*}
        f\succsim g \iff \alpha f + (1 - \alpha )h \succsim \alpha g + (1 - \alpha )h. 
    \end{equation*}
\end{axiom}

We refer to the set of these axioms (i.e.,  \textit{non-triviality}, \textit{continuity}, and the above four ones)  as \textbf{\textit{axioms of objective rationality}}. \citet{lehrer2011justifiable} proved that a preference  that satisfies \textit{axioms of objective rationality} admits a representation that can be viewed as a generalization of the Bewley model (\citet{bewley2002knightian}) and the justifiable model (Theorem 1 of \citet{lehrer2011justifiable}). 
The formal definition is as follows: 

\begin{definition}
\label{def_genBew}
    For a nonconstant affine function $u:X \to \mathbb{R}$ and a belief collection $\mathbb{P}$, a binary relation $\succsim$ over $\mathcal{F}$ admits a \textit{\textbf{generalized Bewley representation}} $(u, \mathbb{P})$ if for all $f, g\in \mathcal{F}$, 
    \begin{equation}
    \label{eq:genBewley}
        f\succsim g \iff \max_{P\in \mathbb{P}} \min_{p\in P} \qty{  \mathbb{E}_p [u(f)] -  \mathbb{E}_p [u(g)] } \geq 0. 
    \end{equation}
\end{definition}

A preference with a generalized Bewley representation $(u,\mathbb{P})$ evaluates $f$ to be better than $g$ if and only if there exists some prior set $P\in \mathbb{P}$ such that for any $p\in P$, the expected utility of $f$ is higher than that of $g$. 
For instance, each $P \in \mathbb{P}$ can be interpreted as some theoretical structure in the DM's mind. 
Each of them makes consistent suggestions to the DM but sometimes says nothing since comparisons of some pairs of acts are beyond the scope. 
Furthermore, disagreements sometimes exist among theoretical structures, leading to violations of \textit{transitivity}.

Note that in \eqref{eq:genBewley}, if $\mathbb{P}$ is a singleton, then it becomes a Bewley preference.
Moreover, if each $P \in \mathbb{P}$ is a singleton, then it becomes a justifiable preference. 
At the axiomatic level, the generalized Bewley representations can be characterized by the common axioms used in the characterizations of the Bewley preferences and the justifiable preferences.

\begin{remark}
    In Definition \ref{def_genBew}, $\mathbb{P}$ is a belief collection, that is, a compact set. 
    Strictly speaking, \citet{lehrer2011justifiable} did not show that $\mathbb{P}$ is compact: Instead, they showed that it is ``loosely closed" (see Appendix for a formal definition). 
    To address this gap, we provide proof that we can always choose a compact collection $\mathbb{P}$ that represents any preference that satisfies \textit{axioms of objective rationality}. 
\end{remark}

\begin{remark}
    Focusing on $\succsim^\ast$ that admits a generalized Bewley representation is not a restrictive assumption.
    As shown by \cite{nishimura2016utility}, in a more general setup, any reflexive preference can be represented in a similar way using a collection of sets of utility functions.  
    The generalized Bewley preferences can be considered one of the most straightforward specifications of \citeauthor{nishimura2016utility}'s representations in decision making under uncertainty. Furthermore, a similar model under risk is also known (cf. \citet{hara2019coalitional}). 
\end{remark}

\subsection{The second criterion and axioms for the relationship}

The second binary relation $\succsim^\land$ represents the actual behavior of the DM with $\succsim^\ast$ in mind. 
We assume that $\succsim^\land$ may violate \textit{independence} but satisfies \textit{completeness} and \textit{transitivity}. 
Thus, $\succsim^\land$ can be considered the choice pattern when the DM is compelled to make decisions and behave consistently. 
By imposing axioms about the relationship between $\succsim^\ast$ and $\succsim^\land$, we specify the admissible second criterion. 

We then introduce two axioms about the relationship between them. We introduce the counterpart of \textit{consistency} in GMMS. 
Before introducing the formal definition of our axiom, we point out two drawbacks of using $\succsim^\ast$ directly to construct $\succsim^\land$. 

The first one is that $\succsim^\ast$ is not transitive in general. If we require $\succsim^\land$ to respect $\succsim^\ast$ in any comparison, the obtained second criterion becomes intransitive. 
To address this concern, we focus on the strict part $\succ^\ast$ of $\succsim^\ast$. 
This does not cause any problem of consistency since generalized Bewley preferences satisfy quasi-transitivity. 

The other concern is that $\succsim^\ast$ and $\succ^\ast$ are sensitive to small changes in alternatives. 
Observed data in the real world often contains noise. 
If the ranking among acts is changed when considering small noise, then these evaluations are not reliable. 
\citet{cerreia2020rational} also considered this problem  and introduced a subrelation that is robust to these small perturbations.
For all $f, g\in \mathcal{F}$, we say that $f$ is \textit{robustly better} than $g$ with respect to $\succsim^\ast$, denoted by $f\stsucc^\ast g$, if for all $h , h' \in \mathcal{F}$, there exists $\delta > 0$ such that for all $\varepsilon \in (0, \delta)$, 
\begin{equation*}
    (1 - \varepsilon) f + \varepsilon h \succ^\ast (1 - \varepsilon) g + \varepsilon h'.
\end{equation*}
The relation $f\stsucc^\ast g$ means that $f$ is better than $g$ even if the DM accounts for small misspecifications of data. 
We use this relation to formalize our consistency axiom. 

The first axiom requires that for any two acts $f,g\in \mathcal{F}$, if there exists a constant act $x\in X$ such that $f$ is robustly better than $x$ and $x$ is robustly better than $g$ in the first criterion $\succsim^\ast$, then the second criterion $\succsim^\land$ should conclude that $f$ is strictly better than $g$.

\begin{axiomB}[Robustly Strict Consistency]
    For all $f, g \in \mathcal{F}$, if $f\stsucc^\ast x\stsucc^\ast g$ for some $x\in X$, then $f\succ^\land g$. 
\end{axiomB}

Since we consider a DM who fully understands the value of constant acts \textit{ex ante},  $f\stsucc^\ast x$ yields an unambiguous lower bound of the value of $f$ using a constant act $x$.   Similarly, $x\stsucc^\ast g$ yields an unambiguous upper bound of the value of $g$. 
Thus, $f\stsucc^\ast x\stsucc^\ast g$ can be interpreted as indicating that $f$ is obviously better than $g$, and furthermore, this relation is robust to small noise in observed data. 
\textit{Robustly strict consistency} requires the second relation to respect the first one only if this condition is satisfied. 

The second axiom is a minor modification of \textit{default to certainty} in GMMS. 

\begin{axiomB}[Priority to Certainty]
    For all $f\in \mathcal{F}$ and $x\in X$, $f \not\succ^\ast x$ implies $x\succsim^\land f$.  
\end{axiomB}

This axiom requires the DM to prefer a constant act to an ambiguous one  if there is no strong reason to choose the ambiguous one.

\subsection{Characterization and uniqueness}

We then provide the main result of this section. 
By imposing the axioms that we have introduced, we obtain another foundation of the cautious dual-self EU model. 
Furthermore, due to the uniqueness result of belief collections in the first criterion $\succsim^\ast$, the parameters in the cautious dual-self EU model are uniquely determined as well.

To state the uniqueness part, we introduce several notations. For functions $u$ and $u'$ from $X$ to $\mathbb{R}$, we write $u\approx u'$ if there exist $\alpha > 0$ and $\beta \in \mathbb{R}$ such that $u' = \alpha u + \beta$. 
We call $H \subset \Delta (S)$ a closed half-space if $H = \{ p\in \Delta (S) \mid \mathbb{E}_p[\varphi] \geq \lambda \}$ for some $\varphi \in \mathbb{R}^S$ and $\lambda\in \mathbb{R}$. 
For a belief collection $\mathbb{P}$, its \textit{half-space closure}, denoted by $\overline{\mathbb{P}}$, is defined as
\begin{equation*}
    \overline{\mathbb{P}} = \text{cl} \qty{ H\subset \Delta(S) \mid \text{$H$ is a closed half-space and $P\subset H$ for some $P\in \mathbb{P}$} }, 
\end{equation*}
where $\text{cl}$ denotes the topological closure in $\mathcal{K}(\Delta(S))$ under the Hausdorff topology. 
This concept was first introduced in CFIL. They showed that belief collections of the dual-self EU model are unique with respect to their half-space closure. 
The latter part of the next theorem states that belief collections in the cautious dual-self EU model are also unique in the same sense. 

\begin{theorem}
\label{thm:cautious_rat}
    Let  $\succsim^\ast$ and $\succsim^\land$ be binary relations over $\mathcal{F}$. 
    The following statements are equivalent: 
    \begin{enumerate}[(i)]
    \setlength{\itemsep}{0cm}
    \setlength{\parskip}{0cm}
        \item [(i)] $\succsim^\ast$ satisfies \textit{axioms of objective rationality}; $\succsim^\land$ satisfies \textit{weak order} and \textit{continuity}; and the pair $(\succsim^\ast, \succsim^\land)$ satisfies \textit{robustly strict consistency} and \textit{priority to certainty}. 
        \item [(ii)] There exist a nonconstant affine function $u: X\to \mathbb{R}$ and a belief collection $\mathbb{P}$ such that $\succsim^\ast$ admits the generalized Bewley representation $(u, \mathbb{P})$ and $\succsim^\land$ admits the cautious dual-self EU representation $(u, \mathbb{P})$. 
    \end{enumerate}
    Furthermore, if there exists another pair $(u', \mathbb{P}')$ such that $\succsim^\ast$ admits the generalized Bewley representation $(u', \mathbb{P}')$ and $\succsim^\land$ admits the cautious dual-self EU representation $(u', \mathbb{P}')$, then $u\approx u'$ and $\overline{\mathbb{P}} = \overline{\mathbb{P}'}$. 
\end{theorem}

Since the axioms for the pair $(\succsim^\ast, \succsim^\land)$ are normatively compelling, this theorem can be regarded as providing a normative justification for the cautious dual-self EU models  and, consequently, \textit{weak uncertainty aversion}.  According to this result, preferences for complete hedging and the normative requirement that ``uncertain prospects should be evaluated cautiously” can be reduced to the same functional form.

\begin{remark}
    Note that \citet{lehrer2011justifiable} did not provide the uniqueness of belief collections. Thus, the latter part of Theorem \ref{thm:cautious_rat} is also a contribution of this paper. 
    The technique of using half-space closures was introduced in CFIL to obtain the uniqueness result of the parameters in the dual-self EU model.  
    We prove the uniqueness of belief collections in the generalized Bewley model by developing their technique.   
\end{remark}

\section{Discussion}
\label{sec_disc}

Here, we have considered the invariant biseparable preferences that satisfy \textit{weak uncertainty aversion}. 
Our first result (Theorem \ref{thm:cautiousDSEU}) shows that these preferences characterize the cautious dual-self EU model, where the belief for calculating the expected utility of each act is determined through an intrapersonal three-stage belief-selection game. 
This result is based on the characterization result of the invariant biseparable preferences provided by CFIL. 
In the literature, GMM and \citet{amarante2009foundations} also presented representation theorems for these preferences. 
In Theorems \ref{subthm_2b} and \ref{thm_ama_weakua}, we have provided alternative representations of the preferences in Theorem \ref{thm:cautiousDSEU} in line with the results of GMM and \citeauthor{amarante2009foundations}.
Furthermore, we have provided another foundation for the cautious dual-self EU model in Theorem \ref{thm:cautious_rat}. 
This result shows that given an incomplete and/or intransitive first criterion in some class, the cautious dual-self EU model can be obtained by imposing axioms to construct a rational preference from the first criterion. 

To conclude this paper, we examine extensions of cautious dual-self EU representations and the relationship between \textit{weak uncertainty aversion} and a similar axiom introduced by \citet{siniscalchi2009vector}. 

\subsection{Generalization of variational EU representations}

This paper has examined the implication of \textit{weak uncertainty aversion} based on the maxmin EU model (\citet{gilboa1989maxmin}) and the dual-self EU model (CFIL).
As a generalization of the maxmin EU preferences, MMR proposed and axiomatized a model called the variational EU model. This model captures a wider range of  ambiguity perceptions and attitudes by using a function on the set of beliefs. 

We say that $\succsim$ over $\mathcal{F}$ admits a \textit{variational EU representation} if there exist a nonconstant affine function $u:X\to \mathbb{R}$ and a convex and lower semicontinuous function $c:\Delta (S) \to \mathbb{R}_+\cup \{ +\infty \}$ with $\inf_{p\in \Delta (S)} c(p) = 0$ such that $\succsim$ is represented by the function $U: \mathcal{F} \to \mathbb{R}$ defined as for all $f\in \mathcal{F}$, 
\begin{equation*}
    U(f) = \min_{p\in \Delta (S)} \mathbb{E}_p [u(f)]+c(p).
\end{equation*}
MMR showed that by replacing \textit{certainty independence} with a weaker axiom in  \citeauthor{gilboa1989maxmin}'s result, the preferences that admit variational EU representations can be characterized. 
The formal definition of the weak independence axiom is as follows: 
\begin{axiom}[Weak Certainty Independence]
    For all $f,g\in\mathcal{F}$, $x,y\in X$, and $\alpha\in(0,1)$, 
    \begin{equation*}
        \alpha f+(1-\alpha)x\succsim \alpha g+(1-\alpha)x\implies\alpha f+(1-\alpha)y\succsim \alpha g+(1-\alpha)y.
    \end{equation*}
\end{axiom}
Compared with \textit{certainty independence}, \textit{weak certainty independence} fixes the proportion of mixing. 
Due to this restriction, preferences that violate the scale-invariance property become accommodated. For a detailed explanation, see Example 2 of MMR. 

CFIL examined the preferences that satisfy the axioms of MMR except for \textit{uncertainty aversion} and showed that these preferences can be represented in a  way similar to the dual-self EU model (cf. Theorem 3 of CFIL).  
Formally, a binary relation satisfies these axioms if and only if there exist a nonconstant affine function $u:X\to \mathbb{R}$ and a collection $\mathbb{C}$ of convex functions $c: \Delta(S) \to \mathbb{R}\cup \{+ \infty \}$ with $\max_{c\in\mathbb{C}} \min_{p\in \Delta(S)}  c(p) = 0$ such that $\succsim$ is represented by the function $U: \mathcal{F}\to \mathbb{R}$ defined as for all $f\in \mathcal{F}$, 
\begin{equation}
\label{eq:var-dual}
    U(f) = \max_{c\in\mathbb{C}} \min_{p\in \Delta(S)} \mathbb{E}_p [u(f)]+c(p).
\end{equation}
These functions can be interpreted using an intrapersonal belief-selection game as in the dual-self EU model. 
The only difference is that the action set of the first mover is not a belief collection but a collection of functions. 
CFIL called these representations \textit{variational dual-self EU representations}. 

As an analogue of the cautious dual-self EU model, we can consider the dual-scenario version of \eqref{eq:var-dual}. 

\begin{definition}
    For a nonconstant affine function $u:X\to \mathbb{R}$ and a collection $\mathbb{C}$ of convex functions $c: \Delta(S) \to \mathbb{R}\cup \{+ \infty \}$ with $\max_{c\in\mathbb{C}} \min_{p\in \Delta(S)}  c(p) = 0$, 
    a binary relation $\succsim$ over $\mathcal{F}$ admits a \textbf{\textit{variational cautious dual-self EU representation}} $(u, \mathbb{C})$ if $\succsim$ is represented by the function $U: \mathcal{F}\to \mathbb{R}$ defined as for all $f\in \mathcal{F}$, 
\begin{equation*}
    U(f) = \min\qty{\max_{c\in\mathbb{C}} \min_{p\in \Delta(S)} \mathbb{E}_p [u(f)]+c(p), \min_{c\in\mathbb{C}} \max_{p\in \Delta(S)} \mathbb{E}_p [u(f)]-c(p)}.
\end{equation*} 
\end{definition}

From the argument in Sections \ref{sec_main} and \ref{sec_other}, one might think that these preferences would be characterized by replacing \textit{uncertainty aversion} with \textit{weak uncertainty aversion} in the theorem of MMR. 
However, this conjecture does not hold: These preferences do not always satisfy \textit{weak uncertainty aversion}. We provide a counterexample in Appendix. 

By considering a further weaker axiom and a technical axiom of unboundedness instead of \textit{weak uncertainty aversion}, we can obtain a characterization result. 
For $f\in \mathcal{F}$, we define its \textit{complementary act}, denoted by $\bar{f}$, as for all $s,s'\in S$, 
\begin{equation*}
    \frac{1}{2}f(s)+\frac{1}{2}\bar{f}(s)\sim\frac{1}{2}f(s')+\frac{1}{2}\bar{f}(s'), 
\end{equation*}
if it exists. 
Thus, by mixing $f$ and $\bar{f}$ with equal weights, we can obtain a complete hedge. 
For a pair $(f, \bar{f})$ of acts, we say that it is a \textit{complementary pair} if $ \bar{f}$ is a complementary act of $f$. 

\begin{axiom}[Simple Diversification]
    For all complementary pairs $(f, \bar{f})$ with $f\sim \bar{f}$, $\frac{1}{2}f+\frac{1}{2}\bar{f}\succsim f$.
\end{axiom}
\begin{axiom}[Unboundedness]
    There exist $x,y\in X$ such that for all $\alpha\in(0,1)$, there are $z,z'\in X$ satisfying 
    \begin{equation*}
        \alpha z'+(1-\alpha)y\succ x\succ y \succ\alpha z+(1-\alpha)x.
    \end{equation*}
\end{axiom}
\textit{Simple diversification} was introduced in \citet{siniscalchi2009vector}. 
Since this axiom only considers the pairs of indifferent acts such that the half mixture of them is a perfect hedge, it is weaker than \textit{weak uncertainty aversion}. 
\textit{Unboundedness} is a technical axiom that ensures that the range of the affine function $u:X\to \mathbb{R}$ derived from $\succsim$ is $\mathbb{R}$. This axiom was used in MMR as well. 

We then state our characterization result of the preferences that admit variational cautious dual-self EU representations. 

\begin{theorem}
\label{thm:vari-dual}
    A binary relation $\succsim$ over $\mathcal{F}$ satisfies \textit{weak order}, \textit{continuity}, \textit{monotonicity}, \textit{weak certainty independence}, \textit{simple diversification}, and \textit{unboundedness} if and only if $\succsim$ admits a variational cautious dual-self EU representation $(u, \mathbb{C})$ with $u(X) = \mathbb{R}$. 
\end{theorem}

Note that we do not need to impose \textit{non-triviality} since it is implied by \textit{unboundedness}. 

By Theorem \ref{thm:vari-dual}, we can see that the variational cautious dual-self EU model can accommodate the choice pattern illustrated in the reflection example of \citet{machina2009risk}. This example is known to be incompatible with many models of decision making under uncertainty (see \citet{baillon2011ambiguity}).\footnote{
\citet{siniscalchi2009vector} provides an example of a preference that is compatible with the reflection example (see Example 2 in \citet{siniscalchi2009vector}). 
By \citeauthor{siniscalchi2009vector}'s Theorem 1 and Proposition 2, we can see that this preference satisfies all the axioms used in Theorem \ref{thm:vari-dual} of this paper, which implies that the variational cautious dual-self EU model can accommodate the reflection example. 
}

\subsection{Generalization of uncertainty-averse EU representations}

CMMM considered a more general class of preferences that satisfy \textit{uncertainty aversion}.
They imposed the independence axiom restricted to constant acts. 
\begin{axiom}[Risk Independence]
    For all $x,y,z\in X$ and $\alpha \in (0,1)$, 
    \begin{equation*}
        x\succsim y\iff \alpha x + (1 -\alpha)z \succsim  \alpha y + (1 -\alpha)z . 
    \end{equation*}
\end{axiom}
They characterized the class of preferences that satisfy \textit{risk independence} and \citeauthor{gilboa1989maxmin}'s (\citeyear{gilboa1989maxmin}) axioms except for \textit{certainty independence}. 
Their result states that a preference $\succsim$ satisfies these axioms if and only if for some nonconstant affine function $u: X\rightarrow \mathbb{R}$ and some quasi-convex function $G : \mathbb{R}\times \Delta(S)\rightarrow \mathbb{R}$ such that (i) $G$ is increasing with respect to the first element and (ii) $\inf_{p\in\Delta (S)} G(\gamma, p) = \gamma$ for all $\gamma\in \mathbb{R}$, $\succsim$ is represented by
the function $U: \mathcal{F}\rightarrow \mathbb{R}$ defined as for all $f\in \mathcal{F}$, 
\begin{equation}
\label{eq:unavEU}
    U(f) = \inf_{p\in \Delta(S)} G\qty(\mathbb{E}_p [u(f)], p). 
\end{equation} 
These representations are called \textit{uncertainty-averse EU representations}. 

We then introduce the dual-self form of the  uncertainty-averse EU model. 
We say that $\succsim$ over $\mathcal{F}$ admits a \textit{rational dual-self EU representation} if there exist a nonconstant affine function $u : X\rightarrow \mathbb{R}$ and a collection $\mathbb{G}$ of quasi-convex functions $G :\mathbb{R} \times \Delta (S) \rightarrow\mathbb{R}$ that are increasing in their first argument and satisfy $\max_{G\in \mathbb{G}} \inf_{p\in \Delta (S)} G(\gamma, p) = \gamma$ for all $\gamma \in \mathbb{R}$ such that $\succsim$ is represented by the function $U: \mathcal{F}\rightarrow \mathbb{R}$ defined as for all $f\in \mathcal{F}$, 
\begin{equation}
\label{eq:rational-dual}
    U(f) =  \max_{G\in\mathbb{G}} \inf_{p\in \Delta(S)} G( \mathbb{E}_p [u(f)], p). 
\end{equation}
Theorem 4 of CFIL shows that the rational dual-self EU model can be characterized by the axioms of CMMM except for \textit{uncertainty aversion}. 

As in Theorem \ref{thm:vari-dual}, we characterize the dual-scenario version of 
the rational dual-self EU model. 
To provide a formal definition, we first introduce the dual function of $G$ in \eqref{eq:unavEU} and \eqref{eq:rational-dual}. 
Fix a binary relation $\succsim$ over $\mathcal{F}$ and a nonconstant affine function $u: X \rightarrow \mathbb{R}$ that represents $\succsim$ restricted to $X$. 
Suppose that for any $f\in \mathcal{F}$, there exists a complementary act $\bar{f}$. 
For a function $G:\mathbb{R}\times \Delta(S) \to \mathbb{R}$, define $G^\ast:\mathbb{R} \times \Delta(S)\to \mathbb{R}$ as for all $f\in \mathcal{F}$ and $p\in\Delta(S)$, 
 \begin{equation*}
        G^\ast (\mathbb{E}_p [u(f)] , p) = - G(\mathbb{E}_p [u(\bar{f})] , p) +2 u\qty({ f + \bar{f}\over 2}). 
\end{equation*}
This function $G^\ast$ is a dual of $G$ in the sense that it is reflexive, i.e., $G^{\ast\ast} = G$. Indeed, for all $f\in \mathcal{F}$ and $p\in\Delta(S)$, 
\begin{align*}
    G^{\ast\ast}(\mathbb{E}_p [u(f)], p) &= - G^\ast ( \mathbb{E}_p [u(\bar{f})], p) +2 u\qty({ f + \bar{f}\over 2}) \\
    &= - \qty{- G (\mathbb{E}_p [u(f)] , p) +2 u\qty({ f + \bar{f}\over 2}) }   +2  u\qty({ f + \bar{f}\over 2})\\
    &= G (\mathbb{E}_p [u(f)] , p). 
\end{align*}
Using this definition, we define the dual-scenario version of \eqref{eq:rational-dual}. 


\begin{definition}
\label{def:caurat}
    Let $\succsim$ be a binary relation over $\mathcal{F}$ such that for each $f\in \mathcal{F}$, its complementary act $\bar{f}$ exists. 
    For a nonconstant affine function $u : X\rightarrow \mathbb{R}$ and a collection $\mathbb{G}$ of quasiconvex functions $G : \mathbb{R} \times \Delta (S) \rightarrow\mathbb{R}$ that are increasing in their first argument and satisfy $\max_{G\in \mathbb{G}} \inf_{p\in \Delta (S)} G(\gamma, p) = \gamma$ for all $\gamma\in \mathbb{R}$, $\succsim$  admits a \textit{\textbf{rational cautious dual-self EU representation}} $(u, \mathbb{G})$ if $\succsim$ is represented by the function $U:\mathcal{F}\to \mathbb{R}$ defined as for all $f\in \mathcal{F}$, 
    \begin{equation}
    \label{eq:rat-dual}
        U(f) = \min \qty{\max_{G\in\mathbb{G}} \inf_{p\in \Delta(S)} G( \mathbb{E}_p [u(f)], p), ~ \min_{G\in \mathbb{G}} \sup_{p\in \Delta(S)} G^\ast ( \mathbb{E}_p [u(f)], p)}. 
    \end{equation} 
\end{definition}

Compared with the (variational) cautious dual-self EU model, the action set of the first mover in \eqref{eq:rat-dual} becomes a more general set  $\mathbb{G}$. 
Note that the first assumption in Definition \ref{def:caurat} is always satisfied if $\succsim$ satisfies \textit{weak order}, \textit{continuity}, \textit{monotonicity}, \textit{unboundedness}, and \textit{risk independence} (see the argument in the proof of Theorem \ref{thm:vari-dual}). 

We then state our axiomatization result of the rational cautious dual-self EU model. As in Theorem \ref{thm:vari-dual}, they can be characterized using \textit{simple diversification} and \textit{unboundedness}. 

\begin{theorem}
    A binary relation $\succsim$ over $\mathcal{F}$ satisfies \textit{weak order}, \textit{continuity}, \textit{monotonicity}, \textit{risk independence}, \textit{simple diversification}, and \textit{unboundedness} if and only if $\succsim$ admits a rational cautious dual-self EU representation $(u, \mathbb{G})$ with $u(X)  = \mathbb{R}$. 
\end{theorem}

\subsection{Remark on weak uncertainty aversion and simple diversification}

In Sections \ref{sec_main} and \ref{sec_other}, we have imposed \textit{weak uncertainty aversion} to obtain the representation theorems (Theorems \ref{thm:cautiousDSEU}-\ref{thm_ama_weakua}). On the other hand, a weaker axiom, \textit{simple diversification}, is used to characterize the two general models since they do not necessarily satisfy \textit{weak uncertainty aversion}. 

It should be noted that in Theorems \ref{thm:cautiousDSEU}-\ref{thm_ama_weakua}, we can obtain the same representations even if we impose \textit{simple diversification} instead of \textit{weak uncertainty aversion} because the following result holds. 

\begin{prop}
    If a binary relation $\succsim$ over $\mathcal{F}$ is an invariant biseparable preference, then \textit{weak uncertainty aversion} is equivalent to \textit{simple diversification}. 
\end{prop}

This result directly follows from the proof of Theorem \ref{thm:cautiousDSEU} since we only use \textit{simple diversification} to prove the only-if part. 

We use \textit{weak uncertainty aversion} in Theorems \ref{thm:cautiousDSEU}-\ref{thm_ama_weakua} because this axiom reflects the intuition underlying the Ellsberg paradox. 
Furthermore, by using \textit{weak uncertainty aversion} when examining the invariant bispearable preferences, we can highlight the difference at the axiomatic level among the cautious dual-self EU model and its generalizations.

\section*{Appendix}

\subsection*{A1. Proof of Theorem 1}

First, we prove the only-if part. Let $\succsim$ be a binary relation over $\mathcal{F}$ that satisfies all of the axioms in the statement. 
By Theorem 1 of CFIL, there exist a nonconstant affine function $u:X\to \mathbb{R}$ and a belief collection $\mathbb{P}$ such that $\succsim$ is represented by the function $U:\mathcal{F}\to \mathbb{R}$ defined as for all $f\in \mathcal{F}$, 
\begin{equation*}
        U(f) = \max_{P \in \mathbb{P}} \min_{p \in P}\mathbb{E}_p [u(f)]. 
\end{equation*}
Since $u$ is nonconstant and unique up to positive affine transformations, we assume  $[-1,1]\subset u(X)$ without loss of generality.
Let $x_0 \in X$ be such that $u(x_{0})=0$. Such an outcome $x_0$ exists since $u$ is affine and $[-1,1]\subset u(X)$.

Note that it is sufficient to prove that for all $f\in \mathcal{F}$, 
\begin{equation}
    \label{eq:wts_ineq}
        \min_{P \in \mathbb{P}} \max_{p \in P}\mathbb{E}_p [u(f)] \geq \max_{P \in \mathbb{P}} \min_{p \in P}\mathbb{E}_p [u(f)]. 
\end{equation}
Indeed, if \eqref{eq:wts_ineq} holds, then for all $f\in \mathcal{F}$, 
    \begin{equation*}
        U(f) = \max_{P \in \mathbb{P}} \min_{p \in P}\mathbb{E}_p [u(f)] = 
        \min \qty{ \max_{P \in \mathbb{P}} \min_{p\in P} \mathbb{E}_p [u(f)], \min_{P \in \mathbb{P}} \max_{p\in P} \mathbb{E}_p [u(f)] }, 
    \end{equation*}
that is, $\succsim$ admits the cautious dual-self EU representation $(u, \mathbb{P})$. 

Suppose to the contrary that there exists $f\in \mathcal{F}$ such that 
\begin{equation}
\label{eq:contra}
    \min_{P \in \mathbb{P}} \max_{p \in P}\mathbb{E}_p [u(f)] < \max_{P \in \mathbb{P}} \min_{p \in P}\mathbb{E}_p [u(f)]. 
\end{equation}
Without loss of generality, we can assume $u(f)\in[-\frac{1}{3},\frac{1}{3}]^S$.\footnote{ 
For $\varepsilon \in(0,1)$, define $f^\varepsilon\in\mathcal{F}$ as $f^\varepsilon =\varepsilon f+(1-\varepsilon)x_{0}$. Since $u$ is affine,  $u(f^\varepsilon)=\varepsilon u(f)$ and the counterpart of \eqref{eq:contra} holds. 
By taking $\varepsilon \in (0,1)$ small enough, we can set $f^\varepsilon$ as $u(f^\varepsilon )\in[-\frac{1}{3},\frac{1}{3}]^S$.}

Let $g^*\in\mathcal{F}$ be an act such that $\frac{1}{2}u(f)+\frac{1}{2}u(g^*)=\frac{1}{3}$.
For each $s\in S$, since $u(g^*(s))=\frac{2}{3}-u(f(s))\in[\frac{1}{3},1]$, we have $g^* (s) \succsim f (s)$.
By \textit{monotonicity}, $g^*\succsim f$.
Let $g_*\in\mathcal{F}$ be an act such that $\frac{1}{2}u(f)+\frac{1}{2}u(g_*)=-\frac{1}{3}$.
For each $s\in S$, since $u(g_*(s))=-\frac{2}{3}-u(f(s))\in[-1,-\frac{1}{3}]$, we have $ f (s) \succsim g_* (s)$.
By \textit{monotonicity}, $f\succsim g_*$.
By \textit{continuity}, there exists $\alpha\in[0,1]$ such that $\alpha g^*+(1-\alpha)g_*\sim f$.
Since $u$ is affine, for all $s\in S$, 
\begin{align*}
    &\frac{1}{2}u(\alpha g^* (s) +(1-\alpha)g_* (s))+\frac{1}{2}u(f(s))\\
    &=\alpha\qty(\frac{1}{2}u(g^*(s))+\frac{1}{2}u(f(s)))+(1-\alpha)\qty(\frac{1}{2}u(g_*(s))+\frac{1}{2}u(f(s)))\\
    &=\frac{\alpha}{3}+(1-\alpha)\qty(-\frac{1}{3})\\
    &=\frac{2}{3}\alpha-\frac{1}{3}, 
\end{align*}
that is, $\frac{1}{2}u(\alpha g^*+(1-\alpha)g_*)+\frac{1}{2}u(f)$ is a constant vector in $[0,1]^S$. 

Let $g\in\mathcal{F}$ be such that $g=\alpha g^*+(1-\alpha)g_*$ and $x\in X$ be such that $u(x)=\frac{2}{3}\alpha-\frac{1}{3}$.
By construction, $\frac{1}{2}u(f)+\frac{1}{2}u(g)=u(x)$.
Since $u(g)=-u(f)+2u(x)$, \eqref{eq:contra} implies that 
\begin{align*}
        U(g) 
        &= 
       \max_{P \in \mathbb{P}} \min_{p\in P} \mathbb{E}_p [-u(f)+ 2 u(x)]\\
        &= 
       \max_{P \in \mathbb{P}} \min_{p\in P} - \mathbb{E}_p [u( f)] + 2 u(x) \\
        &= -  \min_{P \in \mathbb{P}} \max_{p\in P}  \mathbb{E}_p [u(f)]  + 2u(x) \\
        &>
        - \max_{P \in \mathbb{P}} \min_{p\in P} \mathbb{E}_p [u( f)] + 2 u(x)\\
        &=-U(f)+2u(x).
    \end{align*}
By \textit{weak uncertainty aversion}, we have $u(x) \geq  U(g)=U(f)$. 
Combined with the above inequality, we have
    \begin{equation*}
        u(x)  >  - U(f) + 2 u(x), 
    \end{equation*}
which is equivalent to $U(f) > u(x)$.
This is a contradiction to $u(x) \geq U(f)$. 

\vspace{5mm}
Next, we prove the converse.
For a nonconstant affine function $u: X\rightarrow \mathbb{R}$ and a belief collection $\mathbb{P}$,  let $\succsim$ be a binary relation over $\mathcal{F}$ that admits the cautious dual-self EU representation $(u, \mathbb{P})$. 
We only show that $\succsim$ satisfies \textit{weak uncertainty aversion} since the other axioms are straightforward to verify. 

Let $f,g\in\mathcal{F}$, $x\in X$, and $\alpha \in (0,1)$ be such that $U(f)=U(g)$ and $u(x) = \alpha u(f) + (1-\alpha ) u(g)$. 
It is sufficient to prove $x\succsim f$, i.e., $u(x) \geq U(f)$. 
Note that  $u(x) = \alpha u(f) + (1-\alpha ) u(g)$ can be rewritten as $u(g) = {1\over 1 - \alpha} u(x) - {\alpha \over 1 - \alpha}u(f) $. For notational simplicity, let $\beta = {1\over 1 - \alpha} >1$. Then, $u(g)=\beta u(x)+(1-\beta ) u(f)$ holds. 
Therefore, we have 
\begin{align*}
        U(g)
        &= \min \qty{ \max_{P \in \mathbb{P}} \min_{p\in P} \mathbb{E}_p [\beta u(x)+(1-\beta ) u(f)], \min_{P \in \mathbb{P}} \max_{p\in P}  \mathbb{E}_p [\beta u(x)+(1-\beta ) u(f)] }
        \\
        &=\beta u(x)+\min \qty{ \max_{P \in \mathbb{P}} \min_{p\in P} (1-\beta) \mathbb{E}_p [u(f)], \min_{P \in \mathbb{P}} \max_{p\in P} (1-\beta) \mathbb{E}_p [u( f)] } \\
        &=\beta u(x)+(1-\beta)\max \qty{ \min_{P \in \mathbb{P}} \max_{p\in P} \mathbb{E}_p [u(f)], \max_{P \in \mathbb{P}} \min_{p\in P} \mathbb{E}_p [u( f)] } \\
        &\leq \beta u(x)+(1-\beta)\min \qty{ \min_{P \in \mathbb{P}} \max_{p\in P} \mathbb{E}_p [u(f)], \max_{P \in \mathbb{P}} \min_{p\in P} \mathbb{E}_p [u( f)] } \\
        &=\beta u(x)+(1-\beta)U(f), 
\end{align*}
where the third equality and the inequality follow from $1-\beta<0$.
Since $U(f)=U(g)$,
    \begin{equation*}
        U(f)\leq \beta u(x)+(1-\beta)U(f)
    \end{equation*}
holds, which is equivalent to $U(f)\leq u(x)$.

\subsection*{A2. Proof of Lemma 1}

Let $\succsim$ over $\mathcal{F}$ be an invariant biseparable preference and $f\in \mathcal{F}$. 
Take $g, g'\in \mathcal{F}$ and $\alpha, \alpha' \in (0,1)$ such that  for all $s,s'\in S$, 
\begin{align*}
    \alpha f(s) + (1- \alpha) g (s) &\sim \alpha f(s') + (1- \alpha) g (s'),\\
    \alpha' f(s) + (1- \alpha') g' (s) &\sim \alpha' f(s') + (1- \alpha') g' (s'). 
\end{align*} 
If $u(g) = u(g')$, then it is straightforward to prove that $g\asymp g'$. 
Thus, we consider the case where $u(g) \neq u(g')$.
Let $x,x' \in X$ be such that $u(x) = \alpha u(f) + (1- \alpha) u(g)$ and $u(x') = \alpha' u(f) + (1- \alpha') u(g')$. 
Since $u(g) \neq u(g')$, we have $u(x)\neq u(x')$.
By construction, there exists an intersection point $\varphi \in \mathbb{R}^S$ between the line connecting $u(g)$ and $u(x')$ and the line connecting $u(g')$ and $u(x)$. 
Let $\lambda, \lambda'\in [0,1]$ be such that $\lambda u(g
) + (1-\lambda) u(x') = \lambda' u(g') + (1-\lambda') u(x) = \varphi$. 
Note that since $\succsim^\#$ admits the Bewley representation $(u, P)$ (cf. Proposition 5 of GMM), it satisfies \textit{monotonicity}. 
Thus, we have $\lambda g + (1-\lambda) x' \sim^\ast \lambda' g' + (1-\lambda') x$, which implies that $g\asymp g'$. 

\subsection*{A3. Proof of Theorem 2(a)}

Before providing a proof of Theorem \ref{subthm_2a}, we prove a lemma about a property of crisp acts. 

\begin{lem}
\label{lem_crisp}
    Let a binary relation $\succsim$ over 
    $\mathcal{F}$ be an invariant biseparable preference. 
    For all $f,g\in\mathcal{F}$, if there exists $\alpha\in(0,1)$ such that  $\alpha f(s)+(1-\alpha)g(s)\sim \alpha f(s')+(1-\alpha)g(s')$ for all $s,s'\in S$ and $f$ is crisp, then $g$ is crisp.
\end{lem}

\begin{proof}
    Let $f,g\in\mathcal{F}$ and $\alpha\in(0,1)$. 
    Suppose that 
    \begin{equation}
    \label{eq:lem_assumption}
        \alpha f(s)+(1-\alpha)g(s)\sim \alpha f(s')+(1-\alpha)g(s')
    \end{equation}
    for all $s,s'\in S$ and $f$ is crisp. 
    Note that by Proposition 10(iii) of GMM, for any $h\in \mathcal{F}$, $h$ is crisp if and only if $\mathbb{E}_p [u(h)]=\mathbb{E}_{p'} [u(h)]$ for all $p,p'\in P$, where $P$ is defined in \eqref{eq:unambiguous}.
    To prove that $g$ is crisp, suppose to the contrary that there exist $q,q'\in P$ such that $\mathbb{E}_{q} [u(g)] \neq\mathbb{E}_{q'} [u(g)]$.
    By \eqref{eq:lem_assumption}, $u(\alpha f+(1-\alpha)g)$ is a constant vector in $\mathbb{R}^S$. 
    Let $x\in X$ be such that $u(x) = u (\alpha f+(1-\alpha)g)$. 
    Since $u$ is affine, $u(g) = {1\over 1 -\alpha} u(x) - {\alpha \over 1 -\alpha} u(f) $. 
    By $\mathbb{E}_q [u(g)] \neq\mathbb{E}_{q'} [u(g)]$, 
    \begin{align*}
       \mathbb{E}_q \qty[ {1\over 1 -\alpha} u(x) - {\alpha \over 1 -\alpha} u(f) ]
       \neq
       \mathbb{E}_{q'} \qty[ {1\over 1 -\alpha} u(x) - {\alpha \over 1 -\alpha} u(f) ],
    \end{align*}
    which is equivalent to $\mathbb{E}_q [u(f)] \neq\mathbb{E}_{q'} [u(f)]$. This is a contradiction to that $f$ is crisp. 
\end{proof}

Then, we prove that (i) implies (ii). Suppose that a binary relation $\succsim$ over $\mathcal{F}$ admits a generalized $\alpha$-maxmin EU representation $(u, P, a)$ such that $P$ is not a singleton.
Let $f,g\in\mathcal{F}$, $x\in X$, and $\alpha \in (0,1)$ be such that $f$ is not crisp, $U(f)=U(g)$, and $u(x)= \alpha u(f)+(1-\alpha)u(g)$.
Note that  the last equality can be rewritten as $u(g) = {1\over 1 - \alpha} u(x) - {\alpha \over 1 - \alpha}u(f) $. For notational simplicity, let $\beta = {1\over 1 - \alpha} >1$. Then, $u(g)=\beta u(x)+(1-\beta ) u(f)$ holds. 
Therefore, we have
\begin{align*}
        U(g)
        &=a([g])\min_{p\in P}\mathbb{E}_p [u(g)]+(1-a([g]))\max_{p\in P}\mathbb{E}_p [u(g)] \\
        &=
        a([g])\min_{p\in P} \mathbb{E}_p [\beta u(x)+(1-\beta)u(f)]
        +(1-a([g]))\max_{p\in P}\mathbb{E}_p [\beta u(x)+(1-\beta)u(f)] \\
        &=\beta u(x)+ a([g])\min_{p\in P}\mathbb{E}_p [(1-\beta)u(f)] +(1-a([g]))\max_{p\in P}\mathbb{E}_p [(1-\beta)u(f)] \\
        &=\beta u(x)+ (1-\beta)\qty[a([g])\max_{p\in P}\mathbb{E}_p [u(f)]+(1-a([g]))\min_{p\in P}\mathbb{E}_p [u(f)]],   
\end{align*}
where the last equality follows from $1 - \beta < 0$. 
\textit{Weak uncertainty aversion} implies $u(x) \geq U(g)$. 
Therefore, by $1- \beta < 0$, we have 
\begin{equation}
\label{eq:alpha_c1}
        a([g])\max_{p\in P}\mathbb{E}_p [u(f)] +(1-a([g]))\min_{p\in P}\mathbb{E}_p [u(f)] \geq u(x).
\end{equation}
On the other hand, by \textit{weak uncertainty aversion}, $u(x) \geq U(f)$. which can be rewritten as 
\begin{equation}
    \label{eq:alpha_c2}
        u(x) \geq a([f])\min_{p\in P}\mathbb{E}_p [u(f)] +(1-a([f]))\max_{p\in P}\mathbb{E}_p [u(f)]. 
\end{equation}
By \eqref{eq:alpha_c1} and \eqref{eq:alpha_c2}, 
\begin{align*}
    &a([g])\max_{p\in P}\mathbb{E}_p [u(f)] +(1-a([g]))\min_{p\in P}\mathbb{E}_p [u(f)] \\
    & ~\geq 
    a([f])\min_{p\in P}\mathbb{E}_p [u(f)] +(1-a([f]))\max_{p\in P}\mathbb{E}_p [u(f)] , 
\end{align*}
which can be rewritten as 
\begin{equation}
\label{eq:alpha_c3}
        ( a([f])+a([g])-1 ) \max_{p\in P} \mathbb{E}_p [u(f)] \geq (a([f])+a([g])-1)\min_{p\in P}\mathbb{E}_p [u(f)].
\end{equation}
Since $f$ is not crisp and $P$ is not a singleton, Proposition 10(iii) of GMM implies that  there exist $q,q'\in P$ such that $\mathbb{E}_q [u(f)]\neq \mathbb{E}_{q'} [u(f)]$.
Without loss of generality, we assume that $\mathbb{E}_q [u(f)] > \mathbb{E}_{q'} [u(f)]$. 
Then, $\max_{p\in P}\mathbb{E}_p [u(f)] \geq\mathbb{E}_q [u(f)] > \mathbb{E}_{q'} [u(f)] \geq\min_{p\in P}\mathbb{E}_{p} [u(f)]$.
Thus, \eqref{eq:alpha_c3} is equivalent to $a([f])+a([g])\geq 1$.
By the definition of $[f]^\ast$, $g\in [f]^\ast$. Therefore, $a([f])+a([f]^\ast)\geq 1$.

\vspace{5mm}
Next, we prove the converse. Let $\succsim$ be a binary relation over $\mathcal{F}$ that admits a generalized $\alpha$-maxmin EU representation $(u, P, a)$ with $a([h])+a([h]^\ast)\geq 1$ for all $h\in\mathcal{F}$ such that $h$ is not crisp. 
Let $f,g\in\mathcal{F}$, $x\in X$, and $\alpha \in (0,1)$ be such that $U(f)=U(g)$ and $u(x) = \alpha u(f) + (1-\alpha ) u(g)$. 
We prove that $u(x) \geq U(f)$. 
If $f$ and $g$ are crisp, then by Proposition 10(iii) of GMM, there exists $\gamma \in \mathbb{R}$ such that for all $p\in P$, $\gamma = \mathbb{E}_p[u(f)] = \mathbb{E}_p[u(g)]$. By the definition of generalized $\alpha$-maxmin EU representations, $\gamma = U(f)$. 
Also, we have 
\begin{align*}
    \gamma = \min_{p\in P}\mathbb{E}_p [ \alpha u(f)+(1-\alpha)u(g) ] & =\max_{p\in P}\mathbb{E}_p [\alpha u(f)+(1-\alpha)u(g) ], 
\end{align*}
which implies that $\gamma = U(\alpha f+(1-\alpha)g)$. Therefore, $u(x) = U(\alpha f+(1-\alpha)g) =U(f)$.

Then, suppose that neither $f$ nor $g$ is not crisp. 
By Lemma \ref{lem_crisp}, it is sufficient to consider the case in which both $f$ and $g$ are not crisp.
Note that  $u(x) = \alpha u(f) + (1-\alpha ) u(g)$ can be rewritten as $u(g) = {1\over 1 - \alpha} u(x) - {\alpha \over 1 - \alpha}u(f) $. For notational simplicity, let $\beta = {1\over 1 - \alpha} >1$. Then, $u(g)=\beta u(x)+(1-\beta ) u(f)$ holds. 
By the definition of $[f]^\ast$, $[f]^\ast = [g]$. 
Then, we have
    \begin{align*}
        U(g)
        &=a([f]^\ast)\min_{p\in P}\mathbb{E}_p [ u(g)] + (1-a([f]^\ast))\max_{p\in P}\mathbb{E}_p [ u(g) ]\\
        &=\beta  u(x)+a([f]^\ast)\min_{p\in P}\mathbb{E}_p [ (1-\beta )u(f) ] +(1-a ([f]^\ast))\max_{p\in P} \mathbb{E}_p [ (1-\beta)u(f)]  \\
        &=\beta u(x)+(1-\beta)\qty[a([f]^\ast)\max_{p\in P}\mathbb{E}_p [  u(f)] +(1-a([f]^\ast))\min_{p\in P} \mathbb{E}_p [ u(f)]] \\
        &\leq \beta u(x)+(1-\beta)\qty[(1-a([f]))\max_{p\in P}\mathbb{E}_p [ u(f)]+a([f])\min_{p\in P}\mathbb{E}_p [ u(f)]] \\
        &=\beta u(x)+(1-\beta)U(f), 
    \end{align*}
where the inequality follows from $a([f])+a([f]^\ast) \geq 1$ and $1 - \beta < 0$.
Since $U(f)=U(g)$, we have $u(x)\geq U(f)$.

\subsection*{A4. Proof of Theorem 2(b)}

Let $\succsim$ be a binary relation over $\mathcal{F}$ that satisfies all of the axioms in the statement. By \hyperref[thm:gmm]{GMM's Representation Theorem} and Theorem \ref{subthm_2a},  $\succsim$ admits a generalized $\alpha$-maxmin EU representation $(u, P, a)$ with $a([h])+a([h]^\ast)\geq 1$ for all $h\in\mathcal{F}$ such that $h$ is not crisp.

Take $f\in \mathcal{F}$ arbitrarily. 
First, suppose that $f$ is crisp.  By Proposition 10(iii) of GMM, $\min_{p\in P} \mathbb{E}_p [u(f)] = \max_{p\in P} \mathbb{E}_p [u(f)]$, which implies that
\begin{align*}
    a([f])\min_{p\in P} \mathbb{E}_p [u(f)]+(1-a([f]))\max_{p\in P}\mathbb{E}_p [u(f)] = \min_{p\in P} \mathbb{E}_p [u(f)]
\end{align*}
and 
\begin{align*}
     (1 - a([f]^\ast) ) \min_{p\in P}\mathbb{E}_p [u(f)]+a([f]^\ast )\max_{p\in P}\mathbb{E}_p [u(f)] = \min_{p\in P} \mathbb{E}_p [u(f)]. 
\end{align*}
Therefore, 
\begin{align*}
    U(f) 
    &= a([f])\min_{p\in P} \mathbb{E}_p [u(f)]+(1-a([f]))\max_{p\in P}\mathbb{E}_p [u(f)] \\
    &= \min_{p\in P} \mathbb{E}_p [u(f)] \\
    &= \min\left\{
           \begin{array}{l}
            a([f])\min_{p\in P}\mathbb{E}_p [u(f)]+(1-a([f]))\max_{p\in P}\mathbb{E}_p [u(f)], \\[2mm]
            (1 - a([f]^\ast) ) \min_{p\in P}\mathbb{E}_p [u(f)]+a([f]^\ast )\max_{p\in P}\mathbb{E}_p [u(f)]
           \end{array}
           \right\}.
\end{align*}

Next, consider the case where $f$ is not crisp. 
Since $a([f])+a([f]^\ast)\geq 1$, 
we have 
    \begin{align*}
        &a([f])\min_{p\in P}\mathbb{E}_p [u(f)]+(1-a([f]))\max_{p\in P}\mathbb{E}_p [u(f)] \\
        &~ \leq 
        (1 - a([f]^\ast) ) \min_{p\in P}\mathbb{E}_p [u(f)]+a([f]^\ast) \max_{p\in P}\mathbb{E}_p [u(f)]. 
    \end{align*}
This implies that 
\begin{align*}
    U(f) 
    &= a([f])\min_{p\in P} \mathbb{E}_p [u(f)]+(1-a([f]))\max_{p\in P}\mathbb{E}_p [u(f)] \\
    &= \min\left\{
           \begin{array}{l}
            a([f])\min_{p\in P}\mathbb{E}_p [u(f)]+(1-a([f]))\max_{p\in P}\mathbb{E}_p [u(f)], \\[2mm]
            (1 - a([f]^\ast) ) \min_{p\in P}\mathbb{E}_p [u(f)]+a([f]^\ast )\max_{p\in P}\mathbb{E}_p [u(f)]
           \end{array}
           \right\}.
\end{align*}

\vspace{5mm}

We prove the second part. Suppose that a binary relation $\succsim$ over $\mathcal{F}$ is represented by \eqref{eq:caudual-genal}. 
Let $f,g\in\mathcal{F}$, $x\in X$, and $\alpha \in (0,1)$ be such that $U(f)=U(g)$ and $u(x) = \alpha u(f) + (1-\alpha ) u(g)$. 
We prove that $x\succsim f$. 
Note that  $u(x) = \alpha u(f) + (1-\alpha ) u(g)$ can be rewritten as $u(g) = {1\over 1 - \alpha} u(x) - {\alpha \over 1 - \alpha}u(f) $. For notational simplicity, let $\beta = {1\over 1 - \alpha} >1$. Then, $u(g)=\beta u(x)+(1-\beta ) u(f)$ holds. 
Therefore, we have 
\begin{align*}
        U(g)
        &= \min\left\{
           \begin{array}{l}
            a([g])\min_{p\in P}\mathbb{E}_p [u(g)]+(1-a([g]))\max_{p\in P}\mathbb{E}_p [u(g)], \\[2mm]
            (1 - a([g]^\ast) ) \min_{p\in P}\mathbb{E}_p [u(g)]+a([g]^\ast )\max_{p\in P}\mathbb{E}_p [u(g)]
           \end{array}
           \right\}
        \\
        &=\beta u(x) \\*
        &~~~~~
        + \min\left\{
           \begin{array}{l}
            a([g])\min_{p\in P}(1-\beta) \mathbb{E}_p [u(f)]+(1-a([g]))\max_{p\in P} (1-\beta) \mathbb{E}_p [u(f)], \\[2mm]
            (1 - a([g]^\ast) ) \min_{p\in P}(1-\beta) \mathbb{E}_p [u(f)]+a([g]^\ast )\max_{p\in P}(1-\beta) \mathbb{E}_p [u(f)]
           \end{array}
           \right\}
        \\
        &=\beta u(x)+  (1-\beta)  \max \left\{
           \begin{array}{l}
            a([g])\max_{p\in P}\mathbb{E}_p [u(f)]+(1-a([g]))\min_{p\in P}  \mathbb{E}_p [u(f)], \\[2mm]
            (1 - a([g]^\ast) ) \max_{p\in P} \mathbb{E}_p [u(f)]+a([g]^\ast )\min_{p\in P} \mathbb{E}_p [u(f)]
           \end{array}
           \right\}
        \\
         &\leq \beta u(x)+  (1-\beta)  \min \left\{
           \begin{array}{l}
            a([f]^\ast)\max_{p\in P}\mathbb{E}_p [u(f)]+(1-a([f]^\ast))\min_{p\in P}  \mathbb{E}_p [u(f)], \\[2mm]
            (1 - a([f]) ) \max_{p\in P} \mathbb{E}_p [u(f)]+a([f] )\min_{p\in P} \mathbb{E}_p [u(f)]
           \end{array}
           \right\}.
\end{align*}
where the third equality and the inequality follow from $1-\beta<0$ and $[f] = [g]^\ast$.
By \eqref{eq:caudual-genal}, $\beta u(x)+  (1-\beta) U(f) \geq U(g)$.
Since $U(f)=U(g)$, we have $u(x) \geq U(f)$.

\subsection*{A5.  Proof of Theorem 3}

First, we prove that (i) implies (ii).  
Let a binary relation $\succsim$ over $\mathcal{F}$ be an invariant biseparable preference that satisfies \textit{weak uncertainty aversion}. 
By \hyperref[thm:ama]{Amarante's Representation Theorem}, there exist a nonconstant affine function $u:X\rightarrow \mathbb{R}$ and a capacity $v$ on $2^{\Delta (S)}$ such that $\succsim$ is represented by the function $U: \mathcal{F}\to \mathbb{R}$ defined as for all $f\in\mathcal{F}$,
    \begin{equation}
    \label{eq:amarante1}
        U(f) = \int\kappa_{u(f)} dv. 
    \end{equation}

Let $f, g\in \mathcal{F}$, $x\in X$ and $\alpha \in (0,1)$ be such that $f\sim g$ and $u(x) = \alpha u(f) + (1 -\alpha)u(g)$. By \textit{weak uncertainty aversion}, $u(x) \geq U(f)$.
Since $u(g) = {1\over 1-\alpha}u(x) - {\alpha \over 1 - \alpha}u(f)$, for all $p\in \Delta (S)$, 
\begin{align*}
     \kappa_{u(g)} (p) 
     &= \mathbb{E}_p [ u(g) ] \\
        &= \mathbb{E}_p  \qty[ {1\over 1-\alpha} u(x)  - {\alpha \over 1 - \alpha} u(f) ] \\
        & ={1\over 1-\alpha}u(x)  - {\alpha \over 1 - \alpha} \mathbb{E}_p [ u(f) ] \\
        &=  {1\over 1-\alpha} u(x)  - {\alpha \over 1 - \alpha}  \kappa_{u(f)} (p), 
\end{align*}
that is, $\kappa_{u(g)}  = {1\over 1-\alpha} u(x)  - {\alpha \over 1 - \alpha}  \kappa_{u(f)}$. 
Since $f\sim g$, i.e., $\int \kappa_{u(f)} dv = \int \kappa_{u(g)} dv$, 
\begin{align*}
    U(f) &= \int  \kappa_{u(f)} dv \\
        &= \alpha \int  \kappa_{u(f)} dv + (1-\alpha) \int  \kappa_{u(g)} dv \\
        &= \alpha \int  \kappa_{u(f)} dv +   (1-\alpha) \qty({1\over 1-\alpha} u(x)  + {\alpha \over 1 - \alpha} \int  - \kappa_{u(f)} dv )\\
        &= u(x) + \alpha \qty( \int  \kappa_{u(f)} dv  + \int -  \kappa_{u(f)} dv ). 
\end{align*}
Since $u(x) \geq U(f)$, we obtain $\int  \kappa_{u(f)} dv  + \int -  \kappa_{u(f)} dv \leq 0$. 
Since this holds for any $f\in \mathcal{F}$, $\int \kappa dv + \int - \kappa dv \leq 0$ for all  affine function $\kappa$ on $\Delta (S)$. 

\vspace{5mm}

Then, we prove that (ii) implies (iii). 
For a  nonconstant affine function $u:X\to \mathbb{R}$ and a capacity $v: 2^{\Delta(S)}\to \mathbb{R}$, let $\succsim$ be a binary relation  over $\mathcal{F}$ that is represented by the function $U: \mathcal{F} \to \mathbb{R}$ defined as $U(f) = \int \kappa_{u(f)} dv$ for all $f\in \mathcal{F}$.
Suppose that
\begin{equation}
\label{eq:thm3_ineq_suplin}
    \int \kappa dv + \int - \kappa dv \leq 0
\end{equation}
 for all affine functions $\kappa$ on $\Delta (S)$.

By the definition of the Choquet integral and the dual capacity,  for all affine functions $\kappa$ on $\Delta (S)$, 
\begin{align*}
         \int - \kappa dv 
         &= 
         \int^0_{-\infty} \qty{ v(-\kappa (p) \geq \beta) -1  } d\beta + \int_0^{\infty}  v(-\kappa (p) \geq \beta) d\beta \\
         &= \int_0^{\infty} \qty{ v(\kappa (p) \leq \beta) -1  } d\beta +   \int^0_{-\infty} v(\kappa (p) \leq \beta) d\beta \\
         &= \int_0^{\infty}  - v^\ast (\kappa (p) \geq \beta)   d\beta +   \int^0_{-\infty} \qty{ 1 - v^\ast(\kappa (p) \geq \beta) } d\beta \\
         &=- \int \kappa dv^\ast.
\end{align*}
By the above and \eqref{eq:thm3_ineq_suplin}, for all $f\in \mathcal{F}$, 
\begin{align*}
        \int \kappa_{u(f)} dv
        &\leq  - \int -  \kappa_{u(f)} dv 
        = \int  \kappa_{u(f)} dv^\ast. 
\end{align*}
Therefore, for all $f\in \mathcal{F}$, 
\begin{equation*}
    U(f) =  \int \kappa_{u(f)} dv = \min \qty{  \int \kappa_{u(f)} dv,  \int \kappa_{u(f)} dv^\ast }. 
\end{equation*}

\vspace{5mm}

Finally, we prove that (iii) implies (i). 
For a nonconstant affine function $u:X\to \mathbb{R}$ and a capacity $v: 2^{\Delta(S)}\to \mathbb{R}$, let $\succsim$ be a binary relation over $\mathcal{F}$ that is represented by the function $U: \mathcal{F} \to \mathbb{R}$ defined as for all $f\in \mathcal{F}$, 
\begin{equation*}
    U(f) =  \min \qty{  \int \kappa_{u(f)} dv,  \int \kappa_{u(f)} dv^\ast }. 
\end{equation*}
We only prove that $\succsim$ satisfies \textit{weak uncertainty aversion}. 

Let $f,g\in\mathcal{F}$, $x\in X$, and $\alpha \in (0,1)$ be such that $U(f)=U(g)$ and $u(x) = \alpha u(f) + (1-\alpha ) u(g)$. 
It is sufficient to prove $x\succsim f$, i.e., $u(x) \geq U(f)$. 
Note that  $u(x) = \alpha u(f) + (1-\alpha ) u(g)$ can be rewritten as $u(g) = {1\over 1 - \alpha} u(x) - {\alpha \over 1 - \alpha}u(f) $. For notational simplicity, let $\beta = {1\over 1 - \alpha} >1$. Then, $u(g)=\beta u(x)+(1-\beta ) u(f)$ holds. 
Therefore, 
\begin{align*}
        U(g)
        &= \min \qty{  \int \kappa_{u(g)} dv,  \int \kappa_{u(g)} dv^\ast } \\
        &= \min \qty{ \beta u(x) +   \int (1-\beta) \kappa_{u(f)} dv, \beta u(x) + \int (1-\beta) \kappa_{u(f)} dv^\ast } \\
        &=\beta u(x)+\min \qty{  \int (1-\beta) \kappa_{u(f)} dv , \int (1-\beta) \kappa_{u(f)}  dv^\ast } \\
        &= \beta u(x)+ (1 -\beta) \max \qty{ \int \kappa_{u(f)} dv^\ast , \int  \kappa_{u(f)}  dv } \\
        &\leq \beta u(x)+ (1 -\beta) \min \qty{  \int \kappa_{u(f)} dv^\ast , \int  \kappa_{u(f)}  dv } \\
        &= \beta u(x)+(1-\beta)U(f), 
    \end{align*}
    where the forth equality and the inequality follow from  $1-\beta<0$ and $\int - \kappa dv = -\int \kappa dv^* $ for all affine functions $\kappa$ on $\Delta(S)$. 
    Since $U(f)=U(g)$,
    \begin{equation*}
        U(f)\leq \beta u(x)+(1-\beta)U(f)
    \end{equation*}
    holds, which is equivalent to $U(f)\leq u(x)$.

\subsection*{A6. Proof of Theorem 4}

We omit a proof that (ii) implies (i) since it is straightforward. 
Suppose that (i) holds. We first prove that $\succsim^\ast$ admits a generalized Bewley representation (Appendix \hyperref[subapp_first]{A6.1}), and then prove that $\succsim^\land$ admits the cautious dual-self EU representation where the parameters coincide with the representation of $\succsim^\ast$  (Appendix \hyperref[subapp_second]{A6.2}). 

\subsubsection*{A6.1 The first criterion}
\label{subapp_first}

First, we prove that $\succsim^\ast$ admits a generalized Bewley representation. 

We say that a collection $\mathbb{Q}$ of nonempty subsets of $\Delta(S)$ is \textit{loosely closed} if for every $Q\in \text{cl} \, \mathbb{Q}$, there exists $Q' \in \mathbb{Q}$ such that $Q' \subset Q$. 
By Theorem 3 of \citet{lehrer2011justifiable}, there exist a nonconstant affine function $u:X\to \mathbb{R}$ and a nonempty loosely closed collection $\mathbb{Q}$ of nonempty closed convex subsets of $\Delta(S)$ such that for all $f, g\in \mathcal{F}$, 
\begin{equation}
\label{eq:LT_thm2}
    f\succsim^\ast g \iff \max_{Q\in \mathbb{Q}} \min_{q \in Q} \qty{\mathbb{E}_q [u(f)] -  \mathbb{E}_q [u(g)]} \geq 0. 
\end{equation}
Note that $\mathbb{Q}$ is not necessarily compact. 

For $\varphi \in \mathbb{R}^S$, define $G_\varphi : \mathcal{K}(\Delta(S)) \to \mathbb{R}$ as for all $P\in \mathcal{K}(\Delta(S))$, $G_\varphi (P) = \min_{p\in P} \mathbb{E}_p [\varphi]$.\footnote{Note that the minimum can be achieved because $P \in \mathcal{K} (\Delta(S))$ is closed.} The following  lemma holds.

\begin{lem}
\label{lem:con}
    For each $\varphi\in \mathbb{R}^S$, the function $G_\varphi : \mathcal{K}(\Delta(S)) \to \mathbb{R}$ is a continuous function in the Hausdorff topology. 
\end{lem}

\begin{proof}
First, we introduce another topology in $\mathcal{K}(\Delta(S))$. 
For any finite collection $\{ Q, Q_1 ,Q_2, \cdots, Q_n \}$ of open subsets of $\Delta(S)$,
let $\text{B} (Q ; Q_1 ,Q_2, \cdots, Q_n )$ be the collection of subsets $R \in \mathcal{K} (\Delta(S))$ such that  $R\subset Q$ and $R\cap Q_i \neq \emptyset$ for each $i =1,2, \cdots, n$. 
These sets form a base for the Vietoris topology on $\mathcal{K} (\Delta(S))$. 
Since $\Delta(S)$ endowed with the Euclidian topology is metrizable, Theorem 3.91 of \citet{AB2006Math} implies that the Hausdorff topology coincides with the the Vietoris topology in $\mathcal{K} (\Delta (S))$. 
Therefore, it is sufficient to prove that $G_\varphi$ is continuous in the Vietoris topology.

Take $P \in \mathcal{K}(\Delta(S))$ and $\varepsilon > 0$ arbitrarily. 
Let $M_1, M_2$ be the open sets of $\Delta(S)$ such that 
\begin{align*}
        M_1 &= \{ \mu \in \Delta(S) \mid  | \mathbb{E}_\mu [\varphi]  -  G_\varphi (P) | < \varepsilon \} , 
        \\
        M_2 &= \{\mu \in \Delta(S) \mid \mathbb{E}_\mu [\varphi]  >  G_\varphi (P) -  \varepsilon \}. 
\end{align*}
Since  $\mathbb{E}_{p}[\varphi] \geq G_\varphi (P)$ for all $p \in P$, we have $P\subset M_2 =  M_1\cup M_2$. Furthermore, $P\cap M_1 \neq \emptyset$ and $P\cap M_2 \neq \emptyset$. Thus, 
$P \in \text{B} (M_1 \cup M_2; M_1, M_2)$. 

Let $P' \in \text{B} (M_1 \cup M_2; M_1, M_2)$. 
By the definition of  $\text{B}  (M_1 \cup M_2; M_1, M_2)$, $P'\subset M_1 \cup M_2$, which implies that for all $p'\in P'$, $\mathbb{E}_{p' }[\varphi] -  G_\varphi (P) > - \varepsilon$. 
Since $P'\cap M_1 \neq \emptyset$, there exists $p^\ast \in P'$ such that $- \varepsilon < \mathbb{E}_{p^\ast} [\varphi]  -  G_\varphi (P)  < \varepsilon$. 
Therefore,  $- \varepsilon < \min_{p'\in P'} \mathbb{E}_{p' }[\varphi] -  G_\varphi (P) <\varepsilon$, which can be rewritten as $| G_\varphi (P') - G_\varphi (P)| < \varepsilon$.
Since $ \text{B} (M_1 \cup M_2; M_1, M_2)$ is a open set including  $P$ in the Vietoris topology, $G_\varphi$ is continuous in the Vietoris topology. 
\end{proof}

Let $\mathbb{P} = \text{cl} \, \mathbb{Q}$. 
Since $\mathcal{K} (\Delta (S))$ is a compact space (cf. Theorem 3.88 of \citet{AB2006Math}), $\mathbb{P}$ is also compact. 
Since $G_\varphi$ is continuous for any $\varphi \in \mathbb{R}^S$ (Lemma \ref{lem:con}), we have 
\begin{equation*}
    \max_{P\in \mathbb{P}} \min_{p\in P} \mathbb{E}_p [\varphi] = \max_{P\in \mathbb{P}} G_\varphi (P) = \max_{Q\in \mathbb{Q}} G_\varphi (Q) = \max_{Q\in \mathbb{Q}} \min_{q\in Q} \mathbb{E}_p [\varphi]. 
\end{equation*}
Therefore, by \eqref{eq:LT_thm2}, for all $f, g\in \mathcal{F}$,
\begin{equation*}
    f\succsim^\ast g\iff\max_{P\in \mathbb{P}} \min_{p \in P} \qty{\mathbb{E}_p [u(f)] -  \mathbb{E}_q [u(g)]} \geq 0,
\end{equation*}
that is, $\succsim^\ast$ admits the generalized Bewley representation $(u, \mathbb{P})$.

\subsubsection*{A6.2 Rationalization procedure}
\label{subapp_rat}


We then prove that $\succsim^\land$ admits the cautious dual-self EU representation $(u, \mathbb{P})$. 

By \textit{robustly strict consistency} and \textit{priority to certainty}, for all $x, y\in X$, 
    \begin{equation}
    \label{eq:u_equivalence}
        x\succsim^\land y \iff x\succsim^\ast y \iff u(x) \geq u(y). 
    \end{equation}

Fix $f\in \mathcal{F}$. We prove that there exists $x_f \in X$ such that $f\sim^\land x_f$. 
Let $x^+, x^- \in X$ be such that $x^+ \succ^\land x^-$ and $x^+ \succsim^\land f(s) \succsim^\land x^-$ for all $s\in S$.\footnote{By \textit{non-triviality} of $\succsim^\ast$ and  
\eqref{eq:u_equivalence}, such a pair $(x^+, x^-)$ exists. }
Take $\alpha \in (0,1)$ arbitrarily. 
By \eqref{eq:u_equivalence},  for all  $s\in S$, $x^+ \succ^\ast \alpha f (s) + (1-\alpha) x^- $. 
Furthermore, since $u$ is affine and $S$ is finite, there exits $y \in X$ such that for all $s\in S$, $x^+ \succ^\ast y \succ^\ast \alpha f (s) + (1-\alpha) x^- $.
Since $\succsim^\ast$ admits the generalized Bewley representation $(u, \mathbb{P})$,  we have  $x^+ \stsucc^\ast y \stsucc^\ast \alpha f + (1-\alpha) x^- $.
By \textit{robustly strict consistency}, $x^+ \succ^\land \alpha f + (1-\alpha) x^- $. 
If $f  \succ^\land x^+ (\succ^\land x^-)$, then \textit{continuity} implies that for some $\hat{\alpha} \in (0,1)$, $\hat{\alpha} f + (1-\hat{\alpha}) x^- \succ^\land x^+$, which is a contradiction. 
Therefore, $x^+ \succsim^\land f$. 

To prove that $f \succsim^\land x^-$, take $\alpha' \in (0,1)$ arbitrarily. 
Similarly, there exists $y'\in X$ such that  for all and $s\in S$, $\alpha' f (s) + (1-\alpha') x^+ \succ^\ast y' \succ^\ast x^-$. 
Since $\succsim^\ast$ admits the generalized Bewley representation $(u, \mathbb{P})$,  we have  $\alpha' f + (1-\alpha') x^+ \stsucc^\ast y' \stsucc^\ast x^-$. 
By \textit{robustly strict consistency}, $\alpha' f + (1-\alpha') x^+ \succ^\land x^-$.
If $(x^+ \succ^\land) x^-  \succ^\land  f$, then \textit{continuity} implies that for some $\check{\alpha} \in (0,1)$, $x^- \succ^\land \check{\alpha} f + (1-\check{\alpha}) x^+$, which is a contradiction. 
Therefore, $f \succsim^\land x^-$.


If $f\sim^\land x^+$ or $f\sim^\land x^-$, then it is sufficient to set $x_f$ to be $x^+$ or $x^-$, respectively. If not, then $x^+ \succ^\land f \succ^\land x^-$ holds. Suppose that there is no $\hat{\beta} \in (0,1)$ such that $f \sim^\land \hat{\beta} x^+ + (1 - \hat{\beta} ) x^-$. 
Note that by \eqref{eq:u_equivalence}, for all $\delta, \delta'\in (0,1)$, $\delta > \delta'$ is equivalent to $ \delta x^+ + (1 - \delta ) x^- \succ  \delta' x^+ + (1 - \delta' ) x^-$.
Thus, for some $\beta^\ast\in (0,1)$ such that  either (i) for all $\beta \in (0,1)$, $\beta x^+ + (1 - \beta ) x^- \succ^\land f$ if $\beta\geq \beta^\ast$ and $f \succ^\land \beta x^+ + (1 - \beta ) x^- $ otherwise or (ii) for all $\beta \in (0,1)$, $\beta x^+ + (1 - \beta ) x^- \succ^\land f$ if $\beta >  \beta^\ast$ and $f \succ^\land \beta x^+ + (1 - \beta ) x^- $ otherwise. 
To consider the case (i), let $\delta^\ast \in (0, \beta^\ast)$. By $\beta^\ast x^+ + (1 - \beta^\ast ) x^- \succ^\land f \succ^\land \delta^\ast x^+ + (1 - \delta^\ast ) x^-$ and \textit{continuity}, there exists $\beta' \in (\delta^\ast, \beta^\ast)$ such that $\beta' x^+ + (1 - \beta' ) x^- \succ^\land f$, which is a contradiction to (i). 
Similarly, we can prove that the case (ii) does not hold. 
Therefore, there exists $\hat{\beta} \in (0,1)$ such that $f \sim^\land \hat{\beta} x^+ + (1 - \hat{\beta} ) x^-$. Let $x_f = \hat{\beta} x^+ + (1 - \hat{\beta} ) x^-$. Then, $x_f\sim^\land f$, as required.

Assume that 
\begin{equation*}
         \max_{P \in \mathbb{P}} \min_{p \in P} \qty{ \mathbb{E}_p [u(f)] - u(x_f) } > 0 
         ~~~ \text{and} ~~~
         \min_{P \in \mathbb{P}} \max_{p \in P} \qty{ \mathbb{E}_p [u(f)] - u(x_f) } > 0. 
\end{equation*}
Then there exists $z\in X$ such that 
\begin{equation*}
    \max_{P \in \mathbb{P}} \min_{p \in P}  \mathbb{E}_p [u(f)] > u(z) > u( x_f)
    ~~~ \text{and} ~~~
    \min_{P \in \mathbb{P}} \max_{p \in P}  \mathbb{E}_p [u(f)] > u(z) > u( x_f). 
\end{equation*}
By the definition of $\stsucc^\ast$,  $f\stsucc^\ast z \stsucc^\ast x_f$. By \textit{robustly strict consistency}, we have $f \succ^\land x_f$, which is a contradiction to the definition of $x_f$. 
Therefore, 
\begin{equation*}
    \max_{P \in \mathbb{P}} \min_{p \in P} \qty{ \mathbb{E}_p [u(f)] - u(x_f) } \leq 0 
    ~~~ \text{or} ~~~
    \min_{P \in \mathbb{P}} \max_{p \in P} \qty{ \mathbb{E}_p [u(f)] - u(x_f) } \leq 0, 
\end{equation*}
    which can be rewritten as 
    \begin{equation}
    \label{eq:rat-di-con}
         u(x_f) \geq  \max_{P \in \mathbb{P}} \min_{p \in P} \mathbb{E}_p [u(f)] 
         ~~~ \text{or} ~~~
         u(x_f) \geq  \min_{P \in \mathbb{P}} \max_{p \in P} \mathbb{E}_p [u(f)]. 
    \end{equation}  

    Consider the case where $\max_{P \in \mathbb{P}} \min_{p \in P} \mathbb{E}_p [u(f)] \geq \min_{P \in \mathbb{P}} \max_{p \in P} \mathbb{E}_p [u(f)]$. Then, by \eqref{eq:rat-di-con}, $u(x_f) \geq  \min_{P \in \mathbb{P}} \max_{p \in P} \mathbb{E}_p [u(f)]$ holds. 
    If $u(x_f) > \min_{P \in \mathbb{P}} \max_{p \in P} \mathbb{E}_p [u(f)]$, then  there exists $\gamma \in (0,1)$ such that 
    \begin{equation*}
        u(x_f)  
        > u(\gamma x_f + (1-\gamma) x^-)  
        > \min_{P \in \mathbb{P}} \max_{p \in P} \mathbb{E}_p [u(f)]. 
    \end{equation*} 
    By the first inequality, $x_f\succ^\land \gamma x_f + (1-\gamma) x^-$. By the second inequality, $\gamma x_f + (1-\gamma) x^- \succsim^\ast f$. 
    By \textit{priority to certainty}, $\gamma x_f + (1-\gamma) x^- \succsim^\land f$. By \textit{transitivity}, $x_f \succ^\land f$, which is a contradiction to the definition of $x_f$. 
    In this case, we have $u(x_f) =  \min_{P \in \mathbb{P}} \max_{p \in P} \mathbb{E}_p [u(f)]$.

    Consider the case where $\min_{P \in \mathbb{P}} \max_{p \in P} \mathbb{E}_p [u(f)]\geq \max_{P \in \mathbb{P}} \min_{p \in P} \mathbb{E}_p [u(f)] $. 
    Then, by \eqref{eq:rat-di-con}, $u(x_f) \geq  \max_{P \in \mathbb{P}} \min_{p \in P} \mathbb{E}_p [u(f)]$ holds. 
    If $u(x_f) > \max_{P \in \mathbb{P}} \min_{p \in P} \mathbb{E}_p [u(f)]$, then there exists $\gamma' \in [0,1)$ such that 
    \begin{equation*}
        u(x_f)  
        > u(\gamma' x_f + (1-\gamma') x^-)  
        > \max_{P \in \mathbb{P}} \min_{p \in P} \mathbb{E}_p [u(f)]. 
    \end{equation*} 
    By the first inequality, $x_f\succ^\land \gamma' x_f + (1-\gamma') x^-$. By the second inequality, $f \not\succsim^\ast \gamma x_f + (1-\gamma) x^-$. 
    By \textit{priority to certainty}, $\gamma' x_f + (1-\gamma') x^- \succsim^\land f$. By \textit{transitivity}, $x_f \succ^\land f$, which is a contradiction to the definition of $x_f$. 
    In this case, we have $u(x_f) =  \max_{P \in \mathbb{P}} \min_{p \in P} \mathbb{E}_p [u(f)]$. 

Therefore, $\succsim^\land$ is represented by the function $U: \mathcal{F} \to \mathbb{R}$ defined as for all $f\in \mathcal{F}$, 
\begin{equation*}
    U(f) = u(x_f) = \min \qty{ \max_{P \in \mathbb{P}} \min_{p \in P} \mathbb{E}_p [u(f)] , \min_{P \in \mathbb{P}} \max_{p \in P} \mathbb{E}_p [u(f)]  }.
\end{equation*} 

\subsubsection{A6.3 Uniqueness}

Finally, we prove the uniqueness part. 
The strategy of our proof is based on the proof of Proposition 5 of CFIL. 
The most important difference is that while their proof depends on the fact that a binary relation is represented by some function, our proof uses only the original binary relation. 
We need such modifications because $\succsim^\ast$ does not necessarily satisfy \textit{weak order}.

Let $\succsim^\ast$ be a binary relation over $\mathcal{F}$ that admits a generalized Bewley representation $(u, \mathbb{P})$. 
Given $\succsim^\ast$, define the binary relation $\succsim^\circ$ over $\mathbb{R}^S$ as for all $\varphi, \chi \in \mathbb{R}^S$, 
$\varphi \succsim^\circ \chi$ if there exist $f,g\in \mathcal{F}$ and $(\alpha, \beta) \in \mathbb{R}_+\times \mathbb{R}$ such that $\alpha u(f) + \beta \mathbf{1} = \varphi$, $\alpha u(g) + \beta\mathbf{1} = \chi$ and $f\succsim^\ast g$. 
Note that $\succsim^\circ$ does not change if  some positive affine transformation is applied to $u$.
That is, $\succsim^\circ$ is unique up to positive affine transformation of $u$. 
For $ \varphi\in \mathbb{R}^S$ and $\lambda \in \mathbb{R}$, let $H_{\varphi, \lambda} = \{ p\in \Delta (S) \mid \mathbb{E}_p[\varphi] \geq \lambda  \}$. 

\begin{lem}
\label{lem_uni}
    Let $\succsim^\ast$ be a binary relation over $\mathcal{F}$ that admits a generalized Bewley representation $(u, \mathbb{P})$. 
    Then $\overline{\mathbb{P}} = \text{\textup{cl}} \{ H_{\varphi, \lambda} \mid \varphi \in \mathbb{R}^S, \lambda\in \mathbb{R},  \varphi  \succsim^\circ \lambda \mathbf{1} \}$. 
\end{lem}

\begin{proof}
    Let $\varphi \in \mathbb{R}^S$ and $\lambda \in \mathbb{R}$ be such that $\varphi \succsim^\circ  \lambda \mathbf{1}$. Since $\succsim^\ast$ admits  a generalized Bewley representation $(u, \mathbb{P})$, there exists $P \in \mathbb{P}$ such that $\min_{p\in P} \mathbb{E}_p [\varphi - \lambda \mathbf{1}] \geq 0$.
    Thus, $P\subset H_{\varphi - \lambda \mathbf{1}, 0} = H_{\varphi, \lambda }$, which implies $H_{\varphi, \lambda } \in \overline{\mathbb{P}}$. 
    Since $\overline{\mathbb{P}}$ is a closed collection, we have $\text{cl} \{ H_{\varphi, \lambda} \mid \varphi \in \mathbb{R}^S, ~ \varphi \succsim^\circ \lambda \mathbf{1} \} \subset \overline{\mathbb{P}}$. 

    To prove the converse, let  $\varphi \in \mathbb{R}^S$ and $\lambda \in \mathbb{R}$  be such that for some $P'\in \mathbb{P}$, $P' \subset H_{\varphi, \lambda}$. 
    Then, by the definition of $H_{\varphi, \lambda}$, 
    \begin{equation*}
        \max_{P \in \mathbb{P}} \min_{p\in P} \mathbb{E}_p [\varphi] \geq \min_{p\in P'}  \mathbb{E}_p [\varphi] \geq \min_{p\in  H_{\varphi, \lambda}} \mathbb{E}_p  [\varphi] = \lambda.
    \end{equation*}
    Therefore, we have $\max_{P \in \mathbb{P}} \min_{p\in P} \mathbb{E}_p [\varphi- \lambda \mathbf{1}] \geq 0$, which implies $\varphi \succsim^\circ \lambda \mathbf{1}$. 
    Hence, $\overline{\mathbb{P}} \subset \text{cl} \{ H_{\varphi, \lambda} \mid \varphi \in \mathbb{R}^S, ~ \varphi  \succsim^\circ \lambda \mathbf{1} \} $. 
\end{proof}

Suppose that $\succsim^\ast$ admits two generalized Bewley representations $(u, \mathbb{P})$ and $(u', \mathbb{P}')$. The proof of $u\approx u'$ is standard. By Lemma \ref{lem_uni} and the uniqueness of $\succsim^\circ$, we have $\overline{\mathbb{P}} = \overline{\mathbb{P}'}$. 

Then we prove the converse. 
Let $\succsim^\ast$ be a binary relation over $\mathcal{F}$ that admits generalized Bewley representation $(u, \mathbb{P})$, and consider another generalized Bewley representation $(u', \mathbb{P}')$ such that $u \approx u'$ and $\overline{\mathbb{P}} = \overline{\mathbb{P}'}$. 
Let $\varphi\in \mathbb{R}^S$. 
It is sufficient to show that 
\begin{equation*}
    \max_{P \in \mathbb{P}} \min_{p\in P} \mathbb{E}_p [\varphi]  = \max_{P' \in \mathbb{P}'} \min_{p'\in P'} \mathbb{E}_{p'} [\varphi] . 
\end{equation*}
Define $\lambda^\ast(\varphi)\in \mathbb{R}$ as $\lambda^\ast (\varphi) = \max\{ \lambda\in \mathbb{R} \mid \varphi \succsim^\circ \lambda \mathbf{1}  \}$, that is, by Lemma \ref{lem_uni}, $\lambda^\ast (\varphi) = \max\{ \lambda\in \mathbb{R} \mid H_{\varphi, \lambda} \in \overline{\mathbb{P}}\}$. 
Then there exist sequences $\{ P'_n \}_{n\in \mathbb{N}} \subset \mathbb{P}'$ and $\{ H_n \}_{n\in \mathbb{N}}$ such that for each $n\in \mathbb{N}$, $H_n$ is a half space including $P'_n$ and $H_n \to H_{\varphi, \lambda^\ast(\varphi)}$ as $n\to +\infty$. 
Then,  we have
\begin{equation*}
    \min_{p\in H_{\varphi, \lambda^\ast(\varphi)}} \mathbb{E}_p[\varphi] = \lambda^\ast(\varphi) = \lim_{n\to +\infty } \min_{p\in H_n} \mathbb{E}_p[\varphi]
\end{equation*}
and for all $n\in \mathbb{N}$, 
\begin{equation*}
    \min_{p\in H_n} \mathbb{E}_p[\varphi]  \leq \min_{p\in P'_n} \mathbb{E}_p[\varphi].
\end{equation*}
Since $\{ P'_n \}_{n\in \mathbb{N}} \subset \mathbb{P}'$, we have $\max_{P'\in \mathbb{P}'} \min_{p'\in P'}\mathbb{E}_{p'}[\varphi] \geq \lambda^\ast (\varphi)$. 
If there exists $P'' \in \mathbb{P}'$ such that $\min_{p'\in P''} \mathbb{E}_{p'}[\varphi] - \lambda^\ast (\varphi) \eqqcolon \varepsilon  > 0$, then $P'' \subset H_{\varphi, \lambda^\ast(\varphi) + \varepsilon}$, which implies $H_{\varphi, \lambda^\ast(\varphi) + \varepsilon} \in \overline{\mathbb{P}'}$.
By $\overline{\mathbb{P}} = \overline{\mathbb{P}'}$, $H_{\varphi, \lambda^\ast(\varphi) + \varepsilon} \in \overline{\mathbb{P}}$. 
This is a contradiction to the definition of $\lambda^\ast(\varphi)$. 

Thus, we have $\max_{P'\in \mathbb{P}'} \min_{p'\in P'} \mathbb{E}_{p'}[\varphi] = \lambda^\ast (\varphi)$.
Since the definition of $\lambda^\ast(\varphi)$ implies $\max_{P \in \mathbb{P}} \min_{p\in P} \mathbb{E}_{p}[\varphi]  = \lambda^\ast(\varphi)$, we have $\max_{P \in \mathbb{P}} \min_{p\in P} \mathbb{E}_{p}[\varphi]  = \max_{P' \in \mathbb{P}'} \min_{p'\in P'}\mathbb{E}_{p'}[\varphi]$.

\subsection{A7. Counterexample of Section 5.1}
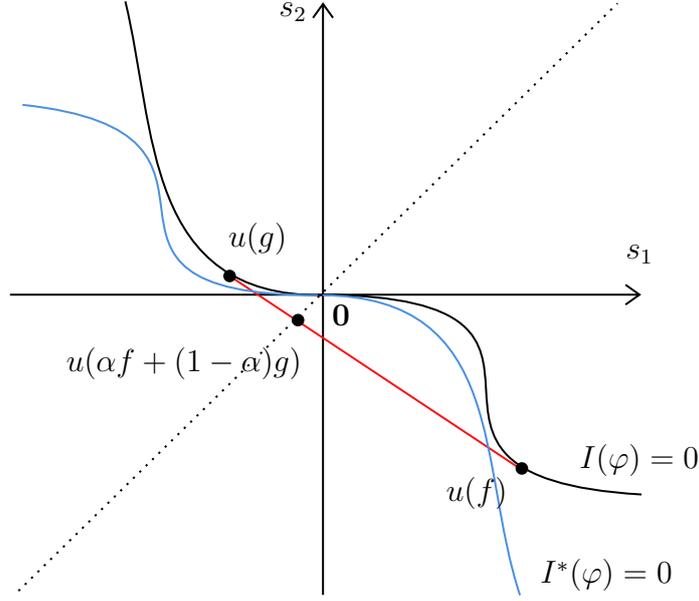
\begin{figure}
    \centering

\tikzset{every picture/.style={line width=0.75pt}} 

\begin{tikzpicture}[x=0.75pt,y=0.75pt,yscale=-1,xscale=1]

\draw  (93.75,199.5) -- (408,199.5)(250,53.25) -- (250,350.5) (401,194.5) -- (408,199.5) -- (401,204.5) (245,60.25) -- (250,53.25) -- (255,60.25)  ;
\draw  [dash pattern={on 0.84pt off 2.51pt}]  (98,348.5) -- (397,53) ;
\draw [color={rgb, 255:red, 253; green, 16; blue, 16 }  ,draw opacity=1 ]   (203.4,190.1) -- (347.25,286.25) ;
\draw  [fill={rgb, 255:red, 0; green, 0; blue, 0 }  ,fill opacity=1 ] (346.6,286.9) .. controls (346.6,285.44) and (347.79,284.25) .. (349.25,284.25) .. controls (350.71,284.25) and (351.9,285.44) .. (351.9,286.9) .. controls (351.9,288.36) and (350.71,289.55) .. (349.25,289.55) .. controls (347.79,289.55) and (346.6,288.36) .. (346.6,286.9) -- cycle ;
\draw  [fill={rgb, 255:red, 0; green, 0; blue, 0 }  ,fill opacity=1 ] (234.88,212.3) .. controls (234.88,210.84) and (236.07,209.65) .. (237.53,209.65) .. controls (239,209.65) and (240.18,210.84) .. (240.18,212.3) .. controls (240.18,213.76) and (239,214.95) .. (237.53,214.95) .. controls (236.07,214.95) and (234.88,213.76) .. (234.88,212.3) -- cycle ;
\draw  [fill={rgb, 255:red, 0; green, 0; blue, 0 }  ,fill opacity=1 ] (200.75,190.1) .. controls (200.75,188.64) and (201.94,187.45) .. (203.4,187.45) .. controls (204.86,187.45) and (206.05,188.64) .. (206.05,190.1) .. controls (206.05,191.56) and (204.86,192.75) .. (203.4,192.75) .. controls (201.94,192.75) and (200.75,191.56) .. (200.75,190.1) -- cycle ;
\draw    (151.33,52) .. controls (166.67,116) and (160.6,200.6) .. (249.75,199.75) ;
\draw [color={rgb, 255:red, 74; green, 144; blue, 226 }  ,draw opacity=1 ]   (100,104) .. controls (230,118) and (106,200) .. (247.35,199.75) ;
\draw    (250,199.5) .. controls (406,201) and (258,291) .. (409,300) ;
\draw [color={rgb, 255:red, 74; green, 144; blue, 226 }  ,draw opacity=1 ]   (249.75,199.75) .. controls (342.33,200.67) and (327,286) .. (348.33,350.67) ;

\draw (253,202.9) node [anchor=north west][inner sep=0.75pt]    {$\boldsymbol{0}$};
\draw (310.03,289.77) node [anchor=north west][inner sep=0.75pt]    {$u( f)$};
\draw (201.3,161.8) node [anchor=north west][inner sep=0.75pt]    {$u( g)$};
\draw (120.33,224.4) node [anchor=north west][inner sep=0.75pt]    {$u( \alpha f+( 1-\alpha ) g)$};
\draw (399.67,173.07) node [anchor=north west][inner sep=0.75pt]    {$s_{1}$};
\draw (376.67,273.07) node [anchor=north west][inner sep=0.75pt]    {$I( \varphi ) =0$};
\draw (357,331.07) node [anchor=north west][inner sep=0.75pt]    {$I^{\ast }( \varphi ) =0$};
\draw (226.67,51.07) node [anchor=north west][inner sep=0.75pt]    {$s_{2}$};

\end{tikzpicture}

    \caption{A counterexample}
    \label{fig:counter}
\end{figure}

We provide an example of a binary relation over $\mathcal{F}$ that admits a variational cautious dual-self EU representation but violates \textit{weak uncertainty aversion}. 
Let $S= \{ s_1, s_2 \}$. 
Define the  monotone constant-additive function $I: \mathbb{R}^S \to \mathbb{R}$ as its indifference curve is drawn in Figure \ref{fig:counter}.\footnote{For a formal definition of constant-additivity, see Footnote \ref{fn_cons}.} 
Then, by the argument in the proof of Theorem 3 of CFIL, there exists a collection $\mathbb{C}$ of convex functions $c: \Delta(S) \to \mathbb{R}\cup \{+ \infty \}$ with $\max_{c\in\mathbb{C}} \min_{p\in \Delta(S)}  c(p) = 0$  such that for all $\varphi \in \mathbb{R}^S$, 
\begin{equation}
\label{eq:vardual_reducedform}
    I(\varphi) = \max_{c\in \mathbb{C}} \min_{p\in \Delta(S)} \mathbb{E}_p [\varphi] + c(p). 
\end{equation}
Let $\succsim$ be a binary relation over $\mathcal{F}$ such that for some $u: X\to \mathbb{R}$ with $u(X) = \mathbb{R}$, $\succsim$ is represented by the function $U: \mathcal{F} \to \mathbb{R}$ defined as for all $f\in \mathcal{F}$, $U(f) = I(u(f))$.

Define the function $I^\ast: \mathbb{R}^S \to \mathbb{R}$ as for all $\varphi\in \mathbb{R}^S$, 
\begin{equation*}
    I^\ast (\varphi) = \min_{c\in \mathbb{C}} \max_{p\in \Delta(S)} \mathbb{E}_p [\varphi] - c(p). 
\end{equation*}
The indifference curve of $I^\ast$ can be obtained by rotating the indifference curve $I$ when its value equals $0$ by 180 degrees about the origin, so it is drawn as the blue curve in Figure \ref{fig:counter}. 
Since the indifference curve of $I^\ast$ is below that of $I$ and these functions are constant-additive, $I^\ast(\varphi)\geq I(\varphi)$ holds for any $\varphi\in \mathbb{R}^S$. 
Therefore, \eqref{eq:vardual_reducedform} can be rewritten as 
\begin{equation*}
     I(\varphi) = \min \qty{ \max_{c\in \mathbb{C}} \min_{p\in \Delta(S)} \mathbb{E}_p [\varphi] + c(p),   \min_{c\in \mathbb{C}} \max_{p\in \Delta(S)} \mathbb{E}_p [\varphi] - c(p) }. 
\end{equation*}
That is, $\succsim$ admits the variational cautious dual-self EU representation $(u, \mathbb{C})$. 

Take $f, g\in \mathcal{F}$ as in Figure \ref{fig:counter}. Since $I(u(f)) = I(u(g)) $, $f\sim g$. Let $\alpha\in (0,1)$ be such that $u(\alpha f + (1-\alpha)g)$ is a constant vector in $\mathbb{R}^S$ as in Figure \ref{fig:counter}. 
By $u(\alpha f + (1-\alpha)g) (s) < 0$ for each $s \in S$,  $f\succ \alpha f + (1-\alpha)g$. 
Therefore, $\succsim$ does not satisfy \textit{weak uncertainty aversion}.

\subsection{A8. Proof of Theorem 5}

First, we prove the only-if part. Let $\succsim$ be a binary relation over $\mathcal{F}$ that satisfies all of the axioms in the statement. 
By Theorem 3 of CFIL, there exist a nonconstant affine function $u:X\to \mathbb{R}$ and a collection $\mathbb{C}$ of convex functions $c: \Delta(S) \to \mathbb{R}\cup \{+ \infty \}$ with $\max_{c\in\mathbb{C}} \min_{p\in \Delta(S)}  c(p) = 0$ such that $\succsim$ is represented by the function $U:\mathcal{F}\to \mathbb{R}$ defined as for all $f\in \mathcal{F}$, 
\begin{equation*}
        U(f)=\max_{c\in\mathbb{C}} \min_{p\in \Delta(S)} \mathbb{E}_p [u(f)]+c(p).
\end{equation*}
By \textit{unboundedness}, $u(X)=\mathbb{R}$.

It is sufficient to prove that for all $f\in \mathcal{F}$, 
\begin{equation*}
        \max_{c\in\mathbb{C}} \min_{p\in \Delta(S)} \mathbb{E}_p [u(f)]+c(p)\leq \min_{c\in\mathbb{C}} \max_{p\in \Delta(S)} \mathbb{E}_p [u(f)]-c(p).
\end{equation*}
Indeed, if this inequality holds, we obtain
\begin{align*}
        U(f)
        &=\max_{c\in\mathbb{C}} \min_{p\in \Delta(S)} \mathbb{E}_p [u(f)]+c(p) \\
        &=\min\qty{\max_{c\in\mathbb{C}} \min_{p\in \Delta(S)} \mathbb{E}_p [u(f)]+c(p), \min_{c\in\mathbb{C}} \max_{p\in \Delta(S)} \mathbb{E}_p [u(f)]-c(p)}.
\end{align*}

Suppose to the contrary that there exists $f\in\mathcal{F}$ such that 
\begin{equation}
\label{eq:contra_var}
    \max_{c\in\mathbb{C}} \min_{p\in \Delta(S)} \mathbb{E}_p [u(f)]+c(p)> \min_{c\in\mathbb{C}} \max_{p\in \Delta(S)} \mathbb{E}_p [u(f)]-c(p). 
\end{equation}
Let $\overline{u} = \max_{s\in S}u(f(s))$ and $\underline{u} = \min_{s\in S}u(f(s))$.\footnote{Since $S$ is finite, $\overline{u}$ and $\underline{u}$ exist.}
Let $g^*\in\mathcal{F}$ be an act such that $\frac{1}{2}u(f)+\frac{1}{2}u(g^*)=\overline{u}$.
For each $s\in S$, since $u(g^*(s))=2\overline{u}-u(f(s)) \geq \overline{u}$, we have $g^* (s) \succsim f (s)$.
By \textit{monotonicity}, $g^*\succsim f$.
Let $g_*\in\mathcal{F}$ be an act such that $\frac{1}{2}u(f)+\frac{1}{2}u(g_*)=\underline{u}$.
For each $s\in S$, since $u(g_*(s))=2\underline{u}-u(f(s))\leq \underline{u}$, we have $ f (s) \succsim g_* (s)$.
By \textit{monotonicity}, $f\succsim g_*$.
By \textit{continuity}, there exists $\alpha\in[0,1]$ such that $\alpha g^*+(1-\alpha)g_*\sim f$.
Notice that since $u$ is affine, for all $s\in S$, 
\begin{align*}
    &\frac{1}{2}u(\alpha g^* (s) +(1-\alpha)g_* (s))+\frac{1}{2}u(f(s))\\
    &=\alpha\qty(\frac{1}{2}u(g^*(s))+\frac{1}{2}u(f(s)))+(1-\alpha)\qty(\frac{1}{2}u(g_*(s))+\frac{1}{2}u(f(s)))\\
    &=\alpha \overline{u} +(1-\alpha) \underline{u}, 
\end{align*}
that is, $\frac{1}{2}u(\alpha g^*+(1-\alpha)g_*)+\frac{1}{2}u(f)$ is a constant vector in $[\underline{u}, \overline{u}]^S$. 

Let $g\in\mathcal{F}$ be such that $g=\alpha g^*+(1-\alpha)g_*$ and $x\in X$ be such that $u(x)=\alpha \overline{u} +(1-\alpha) \underline{u}$.
By construction, $\frac{1}{2}u(f)+\frac{1}{2}u(g)=u(x)$, that is, $(f,g)$ is a complementary pair. 
Since $u(g)=-u(f)+2u(x)$, \eqref{eq:contra_var} implies that
\begin{align*}
        U(g)&=\max_{c\in\mathbb{C}} \min_{p\in \Delta(S)} \mathbb{E}_p [u(g)]+c(p) \\
        &=\max_{c\in\mathbb{C}} \min_{p\in \Delta(S)} \mathbb{E}_p [2u(x)-u(f)]+c(p) \\
        &=2u(x)+\max_{c\in\mathbb{C}} \min_{p\in \Delta(S)} \mathbb{E}_p [-u(f)]+c(p) \\
        &=2u(x)-\qty[\min_{c\in\mathbb{C}} \max_{p\in \Delta(S)} \mathbb{E}_p [u(f)]-c(p)] \\
        &>2u(x)-\qty[\max_{c\in\mathbb{C}} \min_{p\in \Delta(S)} \mathbb{E}_p [u(f)]+c(p)] \\
        &=2u(x)-U(f).
    \end{align*}
    Since $U(f)=U(g)$, this implies $U(f)>u(x)$.
    On the other hand, \textit{simple diversification} implies that $U(f)\leq u(x)$, which is a contradiction.
    
\vspace{5mm}

Next, we prove the converse. 
For a nonconstant affine function $u:X\to \mathbb{R}$ and a collection $\mathbb{C}$ of convex functions $c: \Delta(S) \to \mathbb{R}\cup \{+ \infty \}$ with $\max_{c\in\mathbb{C}} \min_{p\in \Delta(S)}  c(p) = 0$, let $\succsim$ be a binary relation over $\mathcal{F}$ that admits the variational cautious dual-self EU representation $(u, \mathbb{C})$. 
We only show that $\succsim$ satisfies \textit{simple diversification} since the other axioms are straightforward to verify. 

Take any $f,g\in\mathcal{F}$ and $x\in X$ such that $f\sim g$ and $\frac{1}{2}u(f)+\frac{1}{2}u(g)=u(x)$. 
Since $u(g)=-u(f)+2u(x)$, we have
    \begin{align*}
        U(g)&=\min\qty{\max_{c\in\mathbb{C}} \min_{p\in \Delta(S)} \mathbb{E}_p [u(g)]+c(p), \min_{c\in\mathbb{C}} \max_{p\in \Delta(S)} \mathbb{E}_p [u(g)]-c(p)} \\
        &=\min\qty{\max_{c\in\mathbb{C}} \min_{p\in \Delta(S)} \mathbb{E}_p [2u(x)-u(f)]+c(p), \min_{c\in\mathbb{C}} \max_{p\in \Delta(S)} \mathbb{E}_p [2u(x)-u(f)]-c(p)} \\
        &=2u(x)+\min\qty{-\qty[\min_{c\in\mathbb{C}} \max_{p\in \Delta(S)} \mathbb{E}_p [u(f)]-c(p)], -\qty[\max_{c\in\mathbb{C}} \min_{p\in \Delta(S)} \mathbb{E}_p [u(f)]+c(p)]} \\
        &=2u(x)-\max\qty{\min_{c\in\mathbb{C}} \max_{p\in \Delta(S)} \mathbb{E}_p [u(f)]-c(p), \max_{c\in\mathbb{C}} \min_{p\in \Delta(S)} \mathbb{E}_p [u(f)]+c(p)} \\
        &\leq 2u(x)-\min\qty{\min_{c\in\mathbb{C}} \max_{p\in \Delta(S)} \mathbb{E}_p [u(f)]-c(p), \max_{c\in\mathbb{C}} \min_{p\in \Delta(S)} \mathbb{E}_p [u(f)]+c(p)} \\
        &= 2u(x)-U(f).
    \end{align*}
Since $U(f)=U(g)$, we have $u(x)\geq U(f)$.

\subsection{A9. Proof of Theorem 6}

First, we prove the only-if part. Let $\succsim$ be a binary relation over $\mathcal{F}$ that satisfies all of the axioms in the statement. 
By Theorem 4 of CFIL, there exist a nonconstant affine function $u:X\to \mathbb{R}$ and a collection $\mathbb{G}$ of quasi-convex functions $G :\mathbb{R} \times \Delta (S) \rightarrow\mathbb{R}$ that are increasing in their first argument and satisfy $\max_{G\in \mathbb{G}} \inf_{p\in \Delta (S)} G(\gamma, p) = \gamma$ for all $\gamma\in \mathbb{R}$ such that $\succsim$ is represented by the function $U:\mathcal{F}\to \mathbb{R}$ defined as for all $f\in \mathcal{F}$, 
\begin{equation*}
        U(f) = \max_{G\in\mathbb{G}} \inf_{p\in \Delta(S)} G( \mathbb{E}_p [u(f)], p).
\end{equation*}
By \textit{unboundedness}, $u(X)=\mathbb{R}$.

As in the argument in the proofs of Theorems \ref{thm:cautiousDSEU} and \ref{thm:vari-dual}, it is sufficient to prove that for all $f\in \mathcal{F}$, 
\begin{equation*}
    \max_{G\in\mathbb{G}} \inf_{p\in \Delta(S)} G( \mathbb{E}_p [u(f)], p) \leq \min_{G\in \mathbb{G}} \sup_{p\in \Delta(S)} G^\ast ( \mathbb{E}_p [u(f)], p). 
\end{equation*}

Suppose to the contrary that for some $f\in \mathcal{F}$, 
\begin{equation}
\label{eq:contra_ratCDS}
    \max_{G\in\mathbb{G}} \inf_{p\in \Delta(S)} G( \mathbb{E}_p [u(f)], p) > \min_{G\in \mathbb{G}} \sup_{p\in \Delta(S)} G^\ast ( \mathbb{E}_p [u(f)], p). 
\end{equation}
In a way similar to the argument in the proof of Theorem \ref{thm:vari-dual}, we can prove that there exists a complementary act $\bar{f} \in \mathcal{F}$ of $f$ such that $f\sim\bar{f}$. 
Then, by the definition of $G^\ast$, 
\begin{align*}
        \min_{G\in \mathbb{G}} \sup_{p\in \Delta(S)} G^\ast ( \mathbb{E}_p [u(f)], p) 
        &= \min_{G\in \mathbb{G}} \sup_{p\in \Delta(S)} \qty{ - G ( \mathbb{E}_p [u(\bar{f})], p) + 2   u\qty({ f + \bar{f}\over 2})} 
        \\
        &=2 u\qty({ f + \bar{f}\over 2}) - \max_{G\in\mathbb{G}} \inf_{p\in \Delta(S)} G( \mathbb{E}_p [u(\bar{f})], p) .
\end{align*}
By \eqref{eq:contra_ratCDS},  we have $U(f) >   2u \qty({ f + \bar{f}\over 2}) - U(\bar{f})$. Since  $U(f) = U(\bar{f})$, we have $U(f) > 
u \qty({ f + \bar{f}\over 2})$, which is a contradiction to \textit{simple diversification}. 

\vspace{5mm}

Next, we prove the converse. 
For a nonconstant affine function $u:X\to \mathbb{R}$ with $u(X) = \mathbb{R}$ and a collection $\mathbb{G}$ of quasi-convex functions $G :\mathbb{R} \times \Delta (S) \rightarrow\mathbb{R}$ that are increasing in their first argument and satisfy $\max_{G\in \mathbb{G}} \inf_{p\in \Delta (S)} G(\gamma, p) = \gamma$ for all $\gamma\in \mathbb{R}$,  let $\succsim$ be a binary relation over $\mathcal{F}$ that admits the rational cautious dual-self EU representation $(u, \mathbb{G})$. 
We only show that $\succsim$ satisfies \textit{simple diversification} since the other axioms are straightforward to verify. 

Take any $f\in \mathcal{F}$. 
Since $u \qty( {f + \bar{f} \over 2}) $ is a constant vector in $\mathbb{R}^S$, we have 
 \begin{align*}
         &\min \qty{\max_{G\in\mathbb{G}} \inf_{p\in \Delta(S)} G( \mathbb{E}_p [u(f)], p), ~ \min_{G\in \mathbb{G}} \sup_{p\in \Delta(S)} G^\ast ( \mathbb{E}_p [u(f)], p)} 
         \\ 
         &=  \min \qty{\max_{G\in\mathbb{G}} \inf_{p\in \Delta(S)} G( \mathbb{E}_p [u(f)], p), ~ \min_{G\in \mathbb{G}} \sup_{p\in \Delta(S)} \qty[ - G ( \mathbb{E}_p [u(\bar{f})], p) + 2 u \qty( {f + \bar{f} \over 2})] } 
         \\
         &= 
         2 u \qty( {f + \bar{f} \over 2})  \\*
         &~~~~~~
         + \min \qty{\max_{G\in\mathbb{G}} \inf_{p\in \Delta(S)} \qty[ G( \mathbb{E}_p [u(f)], p) - 2 u \qty( {f + \bar{f} \over 2})] , ~ \min_{G\in \mathbb{G}} \sup_{p\in \Delta(S)} - G ( \mathbb{E}_p [u(\bar{f})], p) } 
         \\
         &= 
         2 u \qty( {f + \bar{f} \over 2}) \\
         &~~~~~~
         - \max \qty{\min_{G\in\mathbb{G}} \sup_{p\in \Delta(S)} \qty[ - G( \mathbb{E}_p [u(f)], p) + 2 u \qty( {f + \bar{f} \over 2})] , ~ \max_{G\in \mathbb{G}} \inf_{p\in \Delta(S)}  G ( \mathbb{E}_p [u(\bar{f})], p) } 
         \\
         &=  2 u \qty( {f + \bar{f} \over 2}) - 
         \max \qty{\min_{G\in\mathbb{G}} \sup_{p\in \Delta(S)}  G^\ast ( \mathbb{E}_p [u(\bar{f})], p) , ~ \max_{G\in \mathbb{G}} \inf_{p\in \Delta(S)}  G ( \mathbb{E}_p [u(\bar{f})], p) } 
         \\
         &\leq  2 u \qty( {f + \bar{f} \over 2}) - 
         \min \qty{\min_{G\in\mathbb{G}} \sup_{p\in \Delta(S)}  G^\ast ( \mathbb{E}_p [u(\bar{f})], p) , ~ \max_{G\in \mathbb{G}} \inf_{p\in \Delta(S)}  G ( \mathbb{E}_p [u(\bar{f})], p) }. 
    \end{align*}
Therefore, we have $U(f) \leq 2 u \qty( {f + \bar{f} \over 2}) - U(\bar{f})$. 
Since $U(f) = U(\bar{f})$, we have $ u \qty( {f + \bar{f} \over 2}) \geq U(f)$.


\bibliographystyle{econ}
\bibliography{reference}

\end{document}